\documentclass[balance,upint,subscriptcorrection,varvw,mathalfa=cal=boondoxo,spanish,french,vietnamese,russian,greek,pdf-a,colorlinks]{asmeconf}

\hypersetup{%
	pdfauthor={John H. Lienhard},									  
	pdftitle={ASME Conference Paper LaTeX Template},                  
	pdfkeywords={ASME conference paper, LaTeX template, BibTeX style},
	pdfsubject = {Describes the asmeconf LaTeX template},			  
	pdflicenseurl={https://ctan.org/pkg/asmeconf},
}

\usepackage{caption} 
\usepackage{multirow}
\usepackage{threeparttable}
\usepackage{booktabs}
\usepackage{amsmath}
\usepackage[numbers,sort&compress]{natbib}
\usepackage[dvipsnames]{xcolor}
\usepackage{setspace}
\usepackage{stfloats}
\usepackage{tabularx}
\usepackage{siunitx}
\usepackage{xcolor}
\definecolor{beaublue}{rgb}{0.74, 0.83, 0.9}
\definecolor{darkgray}{rgb}{0.66, 0.66, 0.66}
\definecolor{lavendergray}{rgb}{0.77, 0.76, 0.82}

\begin{document}


\ConfName{Proceedings of the ASME 2024\linebreak Turbomachinery Technical Conference and Exposition}
\ConfAcronym{GT2024}
\ConfDate{June 24--28, 2024} 
\ConfCity{London, England, United Kingdom} 
\PaperNo{GT2024-127455}


\title{Flow and heat transfer in a rotating disc cavity with axial throughflow\\ at high speed conditions} 
 
%
%
%

\SetAuthors{%
	Ruonan Wang\affil{1},
	John W. Chew\affil{1},
        Feng Gao\affil{2},
	Olaf Marxen\affil{1}
	}

\SetAffiliation{1}{Faculty of Engineering and Physical Sciences, University of Surrey, Guildford, GU2 7XH, UK}
\SetAffiliation{2}{Fluid and Acoustic Engineering Laboratory, Aero-Engine Research Institute, Beihang University, Beijing, 102206, P.R. China}


\maketitle





\begin{abstract}
Flow and heat transfer in a compressor rotating disc cavity with axial throughflow is investigated using wall-modelled large-eddy simulations (WMLES). These are compared to measurements from recently published experiments and used to investigate high Reynolds number effects. The simulations use an open-source CFD solver with high parallel efficiency and employ the Boussinesq approximation for centrifugal buoyancy. Kinetic energy effects (characterised by Eckert number) are accounted for by scaling the thermal boundary conditions from static temperature to rotary stagnation temperature. The WMLES shows very encouraging agreement with experiments up to the highest Reynolds number tested, $Re_\phi=3.0\times10^6$. A further simulation at $Re_\phi=10^7$ extends the investigation to an operating condition more representative of aero engine high pressure compressors. The results support the scaling of shroud heat transfer found at lower $Re_\phi$, but disc heat transfer is higher than expected from a simple extrapolation of lower $Re_\phi$ results. This is associated with transition to turbulence in the disc Ekman layers and is consistent with the boundary layer Reynolds numbers at this condition. The introduction of swirl in the axial throughflow, as may occur at engine conditions, could reduce the boundary layer Reynolds numbers and delay the transition.
\end{abstract}


\vspace{1em}

\begin{nomenclature}
\EntryHeading{\bf Roman letters}
\entry{$a$}{Cob inner radius ($\mathrm{m}$)}
\entry{$b$}{Shroud radius ($\mathrm{m}$)}
\entry{$\mathrm{Bo}$}{Buoyancy number, $\mathrm{Ro} / \sqrt{\beta \Delta T_r}$}
\entry{$C_p$}{Specific heat capacity at constant pressure ($\mathrm{J}\ \mathrm{kg}^{-1}\ \mathrm{K}^{-1}$)}
\entry{$d$}{Cavity width ($\mathrm{m}$)}
\entry{$d_h$}{Axial throughflow hydraulic diameter, $2(a-r_s)$ ($\mathrm{m}$)}

\entry{$\mathrm{Ec}$}{Eckert number, $\Omega^2 b^2 / C_p \Delta T$}
\entry{$\mathrm{Gr}$}{Grashof number, $\mathrm{Re}_\phi^2 \beta \Delta T_r$}
\entry{$I$}{Rothalpy, $C_p T + 0.5{\bf u}^2 - 0.5 (\Omega r)^2$ ($\mathrm{J}{\ }\mathrm{kg}^{-1}$)}
\entry{$L$}{Axial length ($\mathrm{m}$)}
\entry{$N$}{Mesh node number}
\entry{$\mathrm{Nu}_b$}{Shroud Nusselt number, $\langle q_b \rangle (d/2) /(\langle \lambda_c \rangle (T_{r,b} - \langle T_{r,c} \rangle))$}
\entry{$\mathrm{Nu}_b^\prime$}{Shroud Nusselt number defined by inlet temperature, $\langle q_b \rangle (d/2) /(\lambda_0 (T_{r,b} - T_{r,0}))$}
\entry{$\mathrm{Nu}_d$}{Disc Nusselt number, $\langle q_d \rangle b /(\lambda_d (T_{r,d} - \langle T_{r,c} \rangle))$}
\entry{$P$}{Static pressure (Pa)}
\entry{$\Pr$}{Prandtl number, $\nu / \alpha$}
\entry{$\Pr_t$}{Turbulent Prandtl number}
\entry{$Q$}{Wall heat transfer ($\mathrm{W}$)}
\entry{$q$}{Wall heat flux ($\mathrm{W}{\ }\mathrm{m}^{-2}$)}
\entry{$r$}{Radius ($\mathrm{m}$)}
\entry{$r_s$}{Shaft radius ($\mathrm{m}$)}
\entry{$r^*$}{Non-dimensional throughflow radius, $(r-r_s)/(a-r_s)$}
\entry{$R$}{Fillet radius ($\mathrm{m}$)}
\entry{$\mathrm{Ra}$}{Rotational Rayleigh number, $ \Pr {\rho_c ^ 2} \Omega^2 b (T_{r,b} - \overline{T}_{r,c}) (d/2)^3 / (\overline{\mu}_c ^2 \overline{T}_{r,c})$}
\entry{$\mathrm{Re}_\phi$}{Rotational Reynolds number, $\Omega b^2 / \nu_0$}
\entry{$\mathrm{Re}_\delta$}{Boundary layer Reynolds number, $u_{\theta,mid} \delta / \nu_{mid}$}
\entry{$\mathrm{Ro}$}{Rossby number, $W/\Omega a$}
\entry{$s$}{Cob width ($\mathrm{m}$)}
\entry{$T$}{Static temperature ($\mathrm{K}$)}
\entry{$T^*$}{Non-dimensional static temperature, $(T-T_0)/(T_b-T_0)$}
\entry{$T_r$}{Rotary stagnation temperature, $I/C_p$ ($\mathrm{K}$)}
\entry{${T_r}^*$}{Non-dimensional rotary stagnation temperature, $(T_r-T_{r,0})/(T_{r,b}-T_{r,0})$}
\entry{$U$}{Velocity component in absolute frame of reference ($\mathrm{m}{\ }\mathrm{s}^{-1}$)}
\entry{$u$}{Velocity component in relative frame of reference ($\mathrm{m}{\ }\mathrm{s}^{-1}$)}
\entry{$u^*$}{Non-dimensional velocity, $u/(\Omega r)$}
\entry{$u_\tau$}{Friction velocity, $\sqrt{\tau / \rho}$ ($\mathrm{m}{\ }\mathrm{s}^{-1}$)}
\entry{$W$}{Bulk velocity at inlet ($\mathrm{m}{\ }\mathrm{s}^{-1}$)}
\entry{$X$}{Non-dimensional estimated core temperature}
\entry{$y^+$}{Non-dimensional wall distance, ${u_\tau} {\Delta y} / {\nu_0}$}

\EntryHeading{\bf Greek letters}
\entry{$\alpha$}{Thermal diffusivity, $\lambda / \rho C_p$ ($\mathrm{m}^2{\ }\mathrm{s}^{-1}$)}
\entry{$\beta$}{Thermal expansion coefficient, $1/T_0$}
\entry{$\delta$}{Laminar Ekman depth, $\sqrt{\nu_0/\Omega}$ (m)}
\entry{$\Delta$}{Laminar Ekman thickness, $\pi \sqrt{\nu_0/\Omega}$ (m)}
\entry{$\Delta T$}{Temperature difference, $T_b - T_0$ ($\mathrm{K}$)}
\entry{$\Delta T_r$}{Rotary stagnation temperature difference, $T_{r,b} - T_{r,0}$ ($\mathrm{K}$)}
\entry{$\Delta y$}{Near wall grid spacing ($\mathrm{m}$)}
\entry{$\Delta Y$}{Normal distance to the wall ($\mathrm{m}$)}
\entry{$\lambda$}{Thermal conductivity ($\mathrm{W}{\ }\mathrm{m}^{-1}{\ }\mathrm{K}^{-1}$)} 
\entry{$\nu$}{Kinematic viscosity ($\mathrm{m}^2{\ }\mathrm{s}^{-1}$)}

\entry{$\rho$}{Density ($\mathrm{kg}{\ }\mathrm{m}^{-3}$)}
\entry{$\tau$}{Wall shear stress ($\mathrm{kg}{\ }\mathrm{m}^{-1}{\ }\mathrm{s}^{-2}$)}
\entry{$\Omega$}{Angular speed ($\mathrm{rad}{\ }\mathrm{s}^{-1}$)}
\EntryHeading{\bf Superscripts and subscripts}
\entry{$*$}{Non-dimensional variable}
\entry{$0$}{Value at the inlet}
\entry{$a$}{Value on the bore}
\entry{$b$}{Value on the shroud}
\entry{$c$}{Value in the cavity core}
\entry{$d$}{Value on the disc}
\entry{mid}{Value at the mid-axial position}
\entry{rms}{Root mean square of the variable}

\entry{$x,r,\theta$}{Axial, radial, tangential component, respectively}

\EntryHeading{\bf Others}
\entry{$\langle \rangle$}{Average in time and in circumferential direction}
\entry{$\overline{~\cdot~}$}{Average in time and on the surface}

\EntryHeading{\bf Acronyms}
\entry{CFD}{~~~~~~~~~~~~~Computational fluid dynamics}
\entry{LES}{~~~~~~~~~~~~~Large-eddy simulation}
\entry{RANS}{~~~~~~~~~~Reynolds-averaged Navier-Stokes}
\entry{URANS}{~~~~~~~Unsteady RANS}
\entry{WMLES}{~~~~~~Wall-modelled LES}
\entry{WMURANS}{Wall-modelled URANS}

\end{nomenclature}


\section{Introduction}
\label{sec:intro}
High-fidelity numerical modelling of flow and heat transfer in turbomachinery is challenging, especially under high Reynolds number conditions. Rotating disc cavity flow, as found in compressor drums, is particularly challenging to model due to its inherent unsteady and unstable nature. Wall-modelled large-eddy simulation (WMLES), as a compromise between computational accuracy and cost, has been implemented to model this problem in previous studies \cite{wang2023advanced, sun2023parametric, sun2022lowro, gao2022flow, gao2021evaluation}, with very encouraging results. The WMLES shows consistent results but significant speed-up compared to recently published wall-resolved LES studies \cite{pitz2019effect, gao2021ekman, saini2021highro}. Gao and Chew \cite{gao2022flow} applied WMLES to a rotating disc cavity at speeds up to 6000 rpm with a rotational Reynolds number ($\mathrm{Re}_\phi$) up to $2.2\times10^6$. This is relatively high for laboratory experiments but well below some engine representative conditions with $\mathrm{Re}_\phi$ of order $10^7$. Also, although wall models are implemented, the computational cost is still significant (up to $\sim 10^6$ CPU hours) \cite{gao2022flow}, which limits its usage in engineering applications. Accurate and efficient modelling of flow and heat transfer in compressor disc cavities under high $\mathrm{Re}_\phi$ is still challenging, and the flow and heat transfer mechanisms under high $\mathrm{Re}_\phi$ have not been fully explored.

\begin{figure*}[!htbp]
	\centerline{\includegraphics[width=\linewidth]{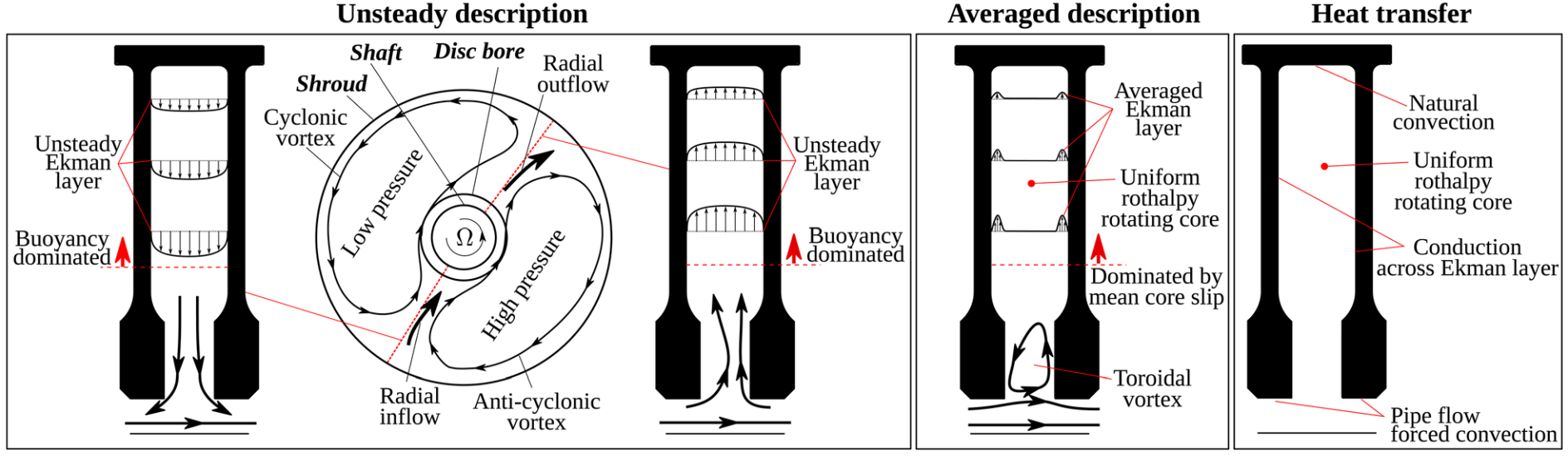}}
	\caption{Summary of flow and heat transfer mechanisms (Gao and Chew, 2022) \cite{gao2022flow}.}
	\label{fig:flow_structure}
\end{figure*}

The investigation of flow and heat transfer in a rotating disc cavity with axial throughflow has been initiated for several decades. A review of the research output in this area to 2015 is reported by Owen and Long \cite{owen2015review}. In recent years, considerable advances have been made through numerical and experimental research. Gao and Chew’s \cite{gao2022flow} summary of the flow and heat transfer mechanisms is given in Fig.~\ref{fig:flow_structure}. This was deduced from studies using WMLES and comparisons with experimental data from the University of Bath \cite{jackson2021analysis, jackson2021measurement} in a combined research programme. The flow away from the discs is turbulent with, in this case, a single cyclonic/anticyclonic vortex pair rotating at a slightly lower speed than the discs. Exchange of flow and energy between the cavity and the axial throughflow and within the cavity occurs principally through the radial inflow and outflow plumes between the vortices. This dominant mixing mechanism is not directly apparent in circumferential and time averages of the flow. The Ekman layers on the discs are highly unsteady, responding to the rotating core flow but remained laminar for the conditions studied. Mean heat transfer from the discs is dominated by conduction across the boundary layer to the rotating core flow with uniform mean rothalpy and rotary stagnation temperature. Shroud heat transfer is dominated by natural convection in the centrifugal force field.

Sun et al.~\cite{sun2023parametric} further examined the heat transfer model indicated in Fig.~\ref{fig:flow_structure} through comparison with WMLES for an aero engine disc cavity geometry with $\mathrm{Re}_\phi=5.4\times10^5$. This is lower than aero engine representative Reynolds numbers, but Rossby number (Ro), buoyancy parameter ($\beta \Delta T_r$), inlet swirl fraction and Eckert number (Ec) were chosen to be representative of engine conditions. An energy balance for the cavity was used to obtain an effective mass flow exchange rate between the axial throughflow and the cavity. Heat transfer on both discs and shroud accounts for the temperature difference between inlet and cavity core. The model was shown to give a useful basis for engineering calculations involving correlation and extrapolation of WMLES and experimental results. Compared to the Bath rig configuration, WMLES on the engine configuration displayed more vortex pairs and shroud Nusselt numbers closer to an established correlation for high Rayleigh number natural convection with $\mathrm{Nu}_b \propto \mathrm{Ra}^{1/3}$.

\begin{figure}[!htbp]
	\centerline{\includegraphics[width=\linewidth]{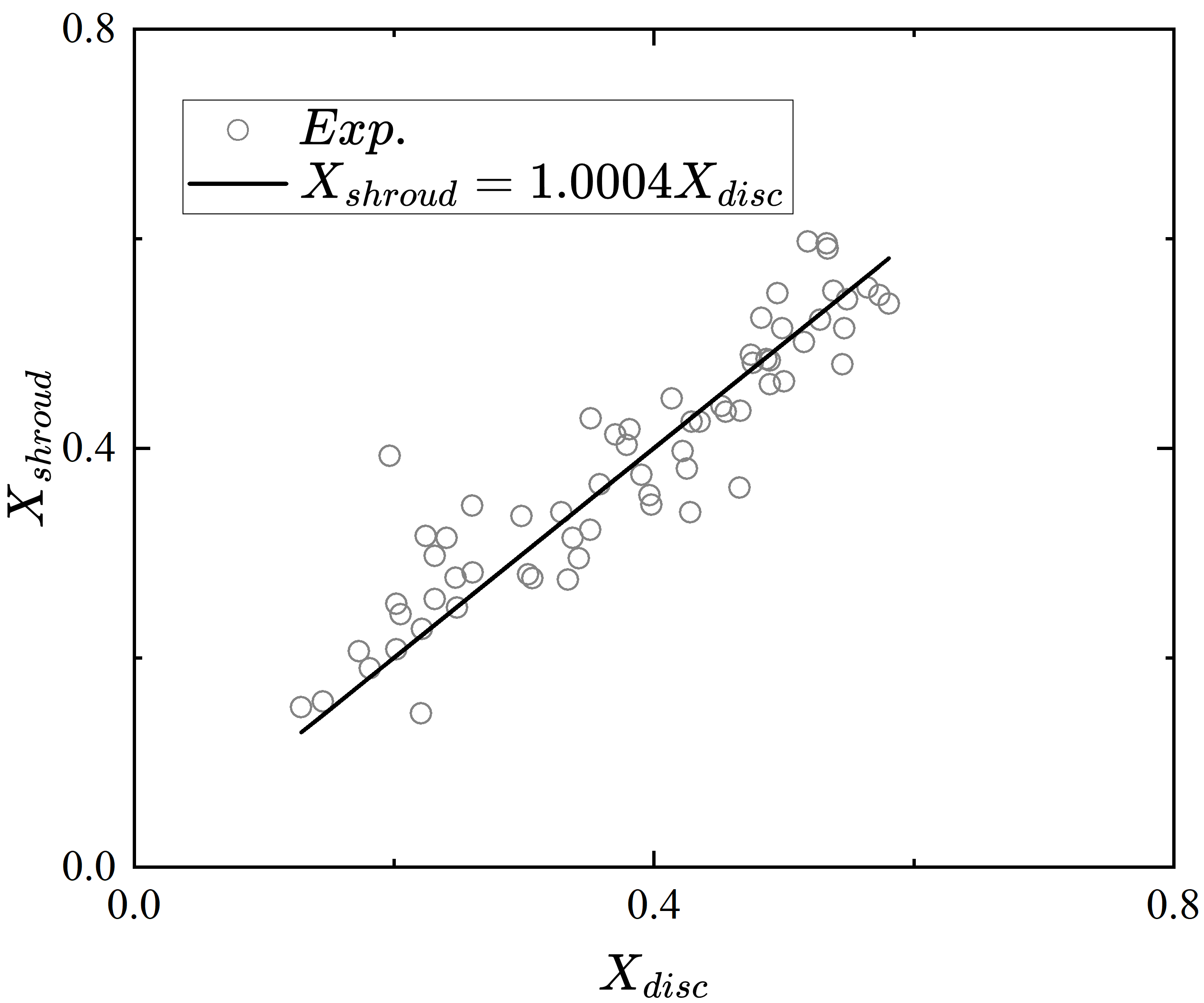}}
	\caption{Estimated nondimensional core temperature from Jackson {et al.’s} \cite{jackson2021analysis, jackson2021measurement} shroud and disc heat transfer measurements.}
	\label{fig:Xshroud_Xdisc}
\end{figure}

Broad agreement of the heat transfer model of Fig.~\ref{fig:flow_structure} with measurements from the University of Bath \cite{jackson2021analysis, jackson2021measurement} is further illustrated in Fig.~\ref{fig:Xshroud_Xdisc}. Each point on the graph corresponds to an experimental condition for which shroud heat transfer measurements and disc heat transfer deduced from temperature measurements were available. For each experimental operating condition, a uniform nondimensional core temperature ($X$), defined in Eq.~(\ref{eq:X}), was estimated independently from either disc or shroud heat transfer. The dimensional core temperatures estimated from disc heat transfer ($T_{r, c, d}$), assuming heat conduction across the disc laminar Ekman layer, are calculated by Eq.~(\ref{eq:Xdisc}). The dimensional core temperatures estimated from shroud heat transfer ($T_{r, c, b}$) assumed $\mathrm{Nu}_b \propto \mathrm{Ra}^{1/3}$. Selecting the constant of proportionality to be 0.61 times the value for the established high Rayleigh number correlation, as in Eq.~(\ref{eq:Xshroud}), gave the results shown with a linear best fit giving $X_{\mathrm{shroud}} \approx X_{\mathrm{disc}}$. This result will be discussed further in comparison with present WMLES results.

\begin{equation}
X=\frac{T_{r,c}-T_{r,0}}{T_{r,b} - T_{r,0}}
\label{eq:X}
\end{equation}

\begin{equation}
\overline{q_d}=\overline{\lambda_d} \frac{\overline{T_{r,d}}-T_{r, c, d}}{\pi \sqrt{\nu / \Omega}}
\label{eq:Xdisc}
\end{equation}

\begin{equation}
\overline{q_b}=\overline{\lambda_b} \frac{0.61 \times 0.15 \times Ra^{1/3} \times (\overline{T_{r,b}}-T_{r, c, b})}{d/2}
\label{eq:Xshroud}
\end{equation}

In high pressure axial compressors, rotational Reynolds numbers may be of order $10^7$ owing to the high rotational speed and high pressure. Matching these conditions in laboratory experiments and large-eddy simulations (LES) is challenging. Hence, although some experiments at $\mathrm{Re}_\phi$ values of around $9\times10^6$ have been achieved at the University of Sussex \cite{long2007shroud, atkins2013inflow} and Technische Universität Dresden \cite{Dresden}, the data available is currently limited. As discussed by Sun et al. \cite{sun2023parametric}, experimental and numerical data at lower $\mathrm{Re}_\phi$ may be scaled to engine conditions, but this increases uncertainties. For example, the scaling of shroud heat transfer with Rayleigh number is still subject to debate and there is potential for transition to turbulence in the disc Ekman layers at high $\mathrm{Re}_\phi$. The present research extends WMLES to higher Reynolds numbers than previously considered. This includes consideration of the mesh requirements, computational cost and parallel performance, and comparison with the highest speed test data available from Jackson et al. \cite{jackson2021analysis, jackson2021measurement}. Following this, a simulation at $\mathrm{Re}_\phi=10^7$ explores flow and heat transfer at a more extreme condition giving new insight.

At high rotational speeds, kinetic energy of the air becomes significant in enthalpy changes when compared to the change due to heat transfer. This is characterised by Eckert number (Ec). Until relatively recently this issue received little attention in the open literature, although there was awareness in industry (that had to consider high speeds) and in related areas. In their discussion of axisymmetric CFD modelling, Kilfoil and Chew \cite{kilfoil2009modelling} considered an “isentropic forced vortex” and assumed instability due to buoyancy when radial gradients of density were lower than that for the isentropic vortex. This criterion can be rewritten as the radial gradient of rotary stagnation temperature being positive. Jackson et al.’s \cite{jackson2021measurement} shroud heat transfer measurements show evidence of Eckert number effects which are described as compressibility effects. In the elementary model described and tested against WMLES with a full compressible solver by Sun et al. \cite{sun2023parametric}, Eckert number effects are captured through the use of rotary stagnation temperature. As will be described below, the present WMLES solves incompressible flow equations with the Boussinesq approximation for buoyancy and use of rotary stagnation temperature rather than static temperature. Further validation of this approach is given through comparison with measurements.    

The rest of the paper is organised as follows. Details of the geometry and operating conditions are described in Section \ref{sec:problems}. Numerical details, including CFD solver, governing equations, model and meshes, convergence conditions, computational cost and parallel performance, are presented in Section \ref{sec:numerical}. Results are given and discussed in Section \ref{sec:results}. Finally, the main conclusions are summarised in Section \ref{sec:conclusions}.

\section{Problem Description}
\label{sec:problems}

\subsection{Geometry}
\label{subsec:geo}

\begin{figure}[!htbp]
  \centerline{\includegraphics[width=\linewidth]{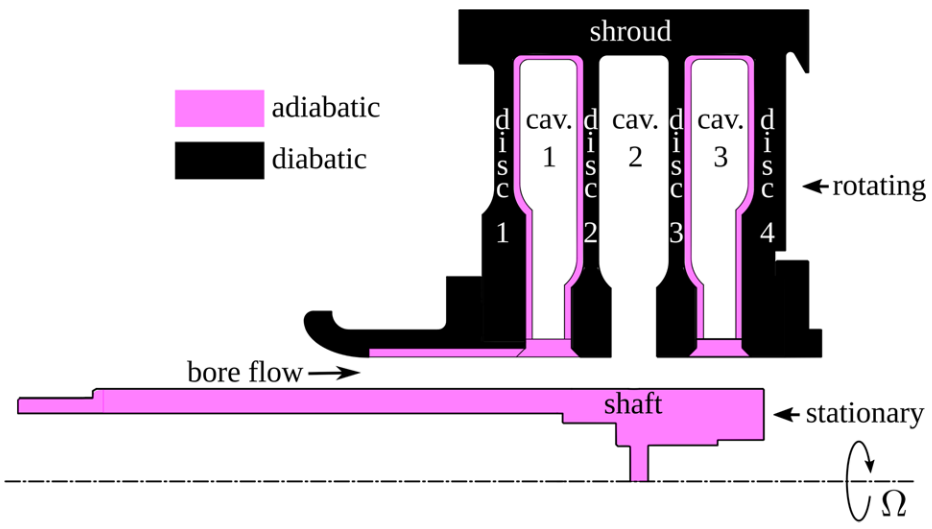}}
  \caption{Rotating disc cavity test rig established at the University of Bath \cite{gao2022flow}.}
  \label{fig:rig_schematic}
\end{figure}

\begin{figure}[!htbp]
  \centerline{\includegraphics[width=\linewidth]{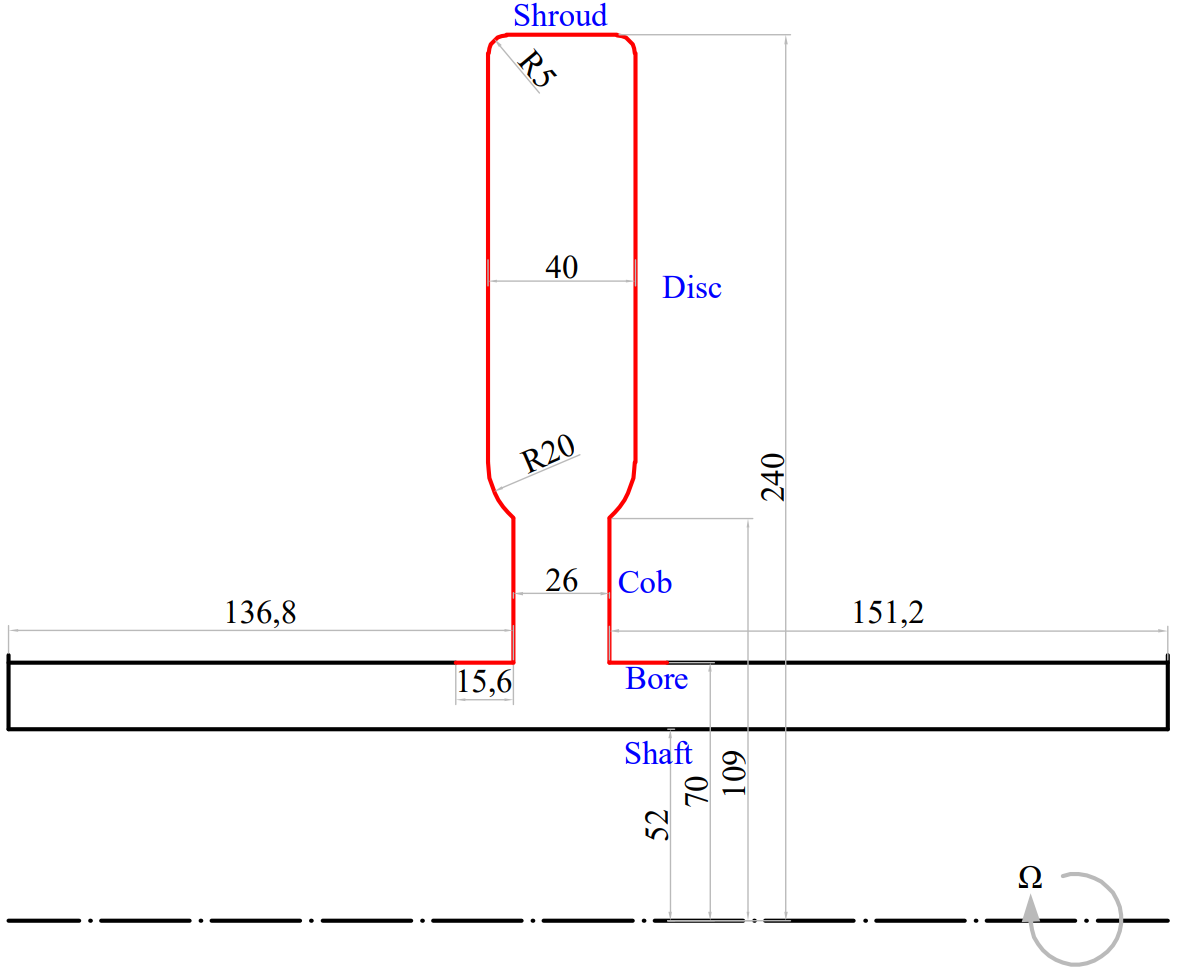}}
  \caption{Schematic and dimensions for the present CFD model [mm].}
  \label{fig:rig_geometry}
\end{figure}

The rotating disc cavity model considers the experimental test rig established at the University of Bath, as shown in Fig.~\ref{fig:rig_schematic}. Only the middle cavity (cav. 2) is open to the bore flow which enters through a bellmouth inlet and exits through radial holes in the rotor downstream of the test section to an extraction unit. A schematic of the present CFD model (corresponding to cav. 2 in Fig.~\ref{fig:rig_schematic}) is shown in Fig.~\ref{fig:rig_geometry}, with dimensions listed in Table \ref{tab:rig_geometry}. More details of the test rig and cavity geometry are given by Luberti et al. \cite{luberti2021design}.

\begin{table*}[!htbp]
	\caption{Dimensions for the rotating disc cavity geometry.}
	\begin{center}
		\label{tab:rig_geometry}
			\begin{tabular}{c c c c c c c c c c c c}
				\toprule
				Parameter & $r_s$ & $a$ & $r_c$ & $b$ & $d$ & $s$ & $L_{bore}$ & $L_{up}$ & $L_{down}$ & $R_b$ & $R_c$ \\
				\midrule
				Dimension [mm] & 52 & 70 & 109 & 240 & 40 & 26 & 15.6 & 136.8 & 151.2 & 5 & 20 \\
				\bottomrule
		\end{tabular}
	\end{center}
\end{table*}

\subsection{Operating Conditions}
\label{subsec:OC}

Operating conditions for the 12 cases simulated are listed in Table~\ref{tab:OC}. Attention is focused on varying Reynolds numbers ($\mathrm{Re}_\phi = 3.2\times10^5-1.0\times10^7$),  buoyancy parameters ($\beta \Delta T_r = 0.14-0.26$) and Rossby numbers ($\mathrm{Ro}=0.1-0.5$). Additionally, two simulations used wall-modelled unsteady Reynolds-averaged Navier-Stokes models (WMURANS) rather than WMLES. For Cases 1 to 11, tags indicate the rotational speed, Rossby number and buoyancy parameter. For Cases 5 and 10, a letter U follows the tag indicating WMURANS. For Case 12, the tag indicates that this has $Re_\phi=10^7$. Air properties for calculating non-dimensional parameters are based on inlet conditions.

\begin{table*}[!htbp]
	\caption{Operating conditions.}
	\begin{center}
		\label{tab:OC}
		\scalebox{1}{
			\begin{tabular}{c c c c c c c c c c c c c c}
				\toprule
				No. & Case Tag & RPM & $\mathrm{Re_\phi}$ & $\mathrm{\beta \Delta T}$ & $\mathrm{\beta \Delta T_r}$ & Ec & Ro & Bo & Gr & Method & Exp. available \\
				\midrule
                1 & 0.8kR5B26              & 800   & $3.2\times10^5$  & 0.26  & 0.26     & 0.005 & 0.5 & 0.99 & $2.7\times10^{10}$ &  WMLES & Yes \\
                2 & 2kR4B26                & 2000  & $8.1\times10^5$  & 0.26  & 0.25     & 0.03  & 0.4 & 0.79 & $1.7\times10^{11}$ &  WMLES & Yes \\
                3 & 6kR2B18                & 6000  & $2.3\times10^6$  & 0.18  & 0.14     & 0.43  & 0.2 & 0.54 & $7.6\times10^{11}$ &  WMLES & Yes \\
				4 & 6kR2B26                & 6000  & $2.3\times10^6$  & 0.26  & 0.20     & 0.31  & 0.2 & 0.44 & $1.1\times10^{12}$ &  WMLES & Yes \\
				5 & 6kR2B26-U              & 6000  & $2.3\times10^6$  & 0.26  & 0.20     & 0.31  & 0.2 & 0.44 & $1.1\times10^{12}$ &  WMURANS & Yes \\   
				6 & 6kR4B18                & 6000  & $2.3\times10^6$  & 0.18  & 0.14   & 0.42 & 0.4 & 1.07 & $7.3\times10^{11}$ &  WMLES & Yes \\
                7 & 6kR4B24                & 6000  & $2.3\times10^6$  & 0.24  & 0.20     & 0.31   & 0.4 & 0.89 & $1.0\times10^{12}$ &  WMLES & Yes \\
	        8 & 8kR1B24                & 8000  & $3.0\times10^6$  & 0.24  & 0.16   & 0.58  & 0.1 & 0.23 & $1.7\times10^{12}$ &  WMLES & No \\	
                9 & 8kR2B24                & 8000  & $3.0\times10^6$  & 0.24  & 0.16   & 0.58  & 0.2 & 0.50 & $1.5\times10^{12}$ &  WMLES & Yes \\
				10 & 8kR2B24-U             & 8000  & $3.0\times10^6$  & 0.24  & 0.16   & 0.58  & 0.2 & 0.50 & $1.5\times10^{12}$ &  WMURANS & Yes \\
		    11 & 8kR4B24               & 8000  & $3.0\times10^6$  & 0.24  & 0.16   & 0.58  & 0.4 & 1.00 & $1.5\times10^{12}$ &  WMLES & No \\
		    12 & Re1E7R4B24            & 8000  & $1.0\times10^7$  & 0.24  & 0.16   & 0.58  & 0.4 & 1.00 & $1.6\times10^{13}$ &  WMLES & No \\                
				\bottomrule
		\end{tabular}}
	\end{center}
\end{table*}

The rotational speeds cover the experimental range from 800 rpm to 8000 rpm. The simulation for Case 1, at 800 rpm, is from a recently published study \cite{wang2023advanced}. For most cases, the operating conditions match experimental conditions. Case 8 is based on Case 9, extending the study to a lower Rossby number by reducing the inlet velocity. Similarly, Case 11 is based on Case 9, increasing the inlet velocity to a higher Rossby number. Case 12 extends Case 11 to a higher Reynolds number by increasing the fluid density (or reducing the kinematic viscosity in the incompressible Navier-Stokes momentum equations) to correspond to a higher pressure experimental condition.

\section{Numerical Details}
\label{sec:numerical}
\subsection{Solver and Governing Equations}
\label{subsec:eqn}

The open-source CFD solver Code\_Saturne is used in the present research. This is a general-purpose CFD solver developed by Électricité de France (EDF) \cite{archambeau2004code}. It is based on a finite volume method (FVM) and centred, second-order scheme for spatial and temporal discretisation. The prospect for large-scale computations was a strong factor in the choice of Code\_Saturne. The parallel performance was previously reported for a local high performance computing (HPC) cluster \cite{wang2023advanced}, and is reported below for simulations on the UK national HPC facility.

The governing equations for conservation of mass, momentum and energy are shown for laminar flow in Eqs. (\ref{eq:continuity}), (\ref{eq:buoyancy}) and (\ref{eq:energy}). These assume a rotating reference frame with reduced pressure calculated by $P^*=P-\rho_0 \Omega^2 r^2/2$. The viscosity term is modified for turbulent flows. The $k-\omega$ SST turbulence model is used for URANS cases. The Smagorinsky subgrid-scale model, with Smagorinsky constant $C_s = 0.065$, is used for WMLES. Laminar viscosity is calculated as a function of temperature using Sutherland's law. Thermal conductivity is calculated as a function of temperature using a quadratic polynomial. Specific heat at constant pressure ($C_p$) is set to a constant 1005 $\mathrm{J}\ \mathrm{kg}^{-1}\ \mathrm{K}^{-1}$. The Prandtl number ($\Pr$) is approximately 0.7 for air, and the turbulent Prandtl number ($\Pr_t$) is set to 0.9.

\begin{equation}
    \nabla \cdot {\bf u}=0
\label{eq:continuity}
\end{equation}

\begin{multline}
\label{eq:buoyancy}
   \rho_{0} \frac{\partial {\bf u}}{\partial t}+\rho_{0} {\bf u} \cdot \nabla {\bf u}= -\nabla {P^*}+\nabla (\mu \nabla {\bf u}) \\\underbrace{-2\rho_{0}{\bf \Omega} \times {\bf u}}_{\text{Coriolis term}} + \underbrace{\rho_{0} \beta \left(T_{0,r}-T_r\right){\bf \Omega} \times ({\bf \Omega} \times {\bf r})}_{\text{Buoyancy generating term}}
\end{multline}

\begin{equation}
   \frac{\partial T_r}{\partial t}+{\bf u} \cdot \nabla T_r = \alpha \nabla^{2} T_r 
\label{eq:energy}
\end{equation}

The simulations follow previous practice \cite{wang2023advanced} with the Boussinesq approximation used to account for buoyancy in the rotating system. This has given good results in comparison to measurements and full compressible flow simulations in previous work \cite{wang2023advanced, wang2023closed} and is retained here for simplicity of implementation.


All the walls rotate at a constant angular speed $\Omega$, except for the stationary shaft. Temperatures are specified on the shroud, discs, cobs and disc bores, shown as red boundaries in Fig.~\ref{fig:rig_geometry}, matching the temperatures measured in the experiments. Other walls are adiabatic. Uniform axial velocity without swirl or turbulence is specified at the inlet matching the Rossby number in the experiment. The inlet swirl effect was considered in Ref. \cite{gao2022flow} by including the swirl generated by the rig bellmouth. In Ref. \cite{sun2022lowro}, the inlet swirl was set to zero, with little effect on the results. Considering this, as well as the inlet swirl being low, the inlet swirl setting in Ref.~\cite{sun2022lowro} was adopted here. Pressure and density correspond to the experimental inlet condition, except for Case 12 in which density is increased to match $\mathrm{Re}_\phi=10^7$.

\subsection{Treatment of Kinetic Energy Effects}
\label{subsec:ke_treatment}
As discussed in Section \ref{sec:intro}, under high speed conditions, the contribution of kinetic energy to enthalpy change is significant and should not be neglected. Here, the kinetic energy effect is taken into account by using the rotary stagnation temperature ($T_r$) in the energy equation (\ref{eq:energy}) and in the Boussinesq approximation. Thus the rotating surface temperatures applied in the simulation are rotary stagnation temperatures calculated from the measured static temperatures and rotational speed. This has enabled efficient calculation for high speed conditions using the incompressible formulation. The formulation neglects frictional heating.

\subsection{Model and Mesh}
\label{subsec:mesh}
All simulations used a full $360^\circ$ CFD model. The mesh details and computational cost for selected cases are listed in Table \ref{tab:mesh}. Mesh resolution is only dependent on $\mathrm{Re}_\phi$ and the meshing criteria adopted. Except for two cases with “Relaxed” criteria, the applied criteria followed those of previous studies \cite{gao2021evaluation, gao2022flow, sun2022lowro, sun2023parametric, wang2023advanced}, with $\Delta x^+ <15$, $\Delta r^+ <40$, $\Delta (r\theta)^+ <30$. The baseline near wall grid spacing is set as the laminar Ekman depth ($\delta$). The time step is set so that the disc travels one grid cell in the circumferential direction per time step, thus the rotational Courant number is equal to one.

\begin{table*}
\caption{Mesh details and computational cost.}
\begin{center}
\label{tab:mesh}
\scalebox{0.95}{
\begin{tabular}{c c c c c c c c c c c c c}
\toprule
No. & $Re_\phi$ & Criteria & $\Delta x^+$ & $\Delta r^+$ & $\Delta (r\theta)^+$ & $\Delta y/\delta$ & $N_x$ & $N_r$ & $N_\theta$ & $N_{total}$ & Revolutions & Cost (CPU hour)\\
\midrule
1        &$3.2\times10^5$ & Standard &  $<15$ &  $<40$ & $<30$ & 1 & 25 & 99 & 400 & $\sim 2.27\times10^6$ & 100+100 & 1194 \\
2        &$8.1\times10^5$ & Standard &  $<15$ &  $<40$ & $<30$ & 1 & 72 & 178 & 600 & $\sim 8.92\times10^6$ & 100+100 & 12012\\
3-7 &$2.3\times10^6$ & Standard &  $<15$ &  $<40$ & $<30$ & 1   & 84& 291 & 800  & $\sim 2.25\times10^7$& 100+100 & 48426\\
7        &$2.3\times10^6$ & Relaxed  &  $<25$ &  $<60$ & $<45$ & 1.5 & 60& 199 & 540  & $\sim 7.28\times10^6$& 100+100 & 9947\\
8-11&$3.0\times10^6$ & Standard &  $<15$ &  $<40$ & $<30$ & 1   & 109 & 226 & 1104 & $\sim 4.18\times10^7$& 100+100 & 121501\\
12       &$1.0\times10^7$ & Relaxed  &  $<25$ &  $<60$ & $<45$ & 1.5 & 156 & 782 & 1600 & $\sim 2.27\times10^8$ & 50+50 & 567457\\
\bottomrule
\end{tabular}}
\end{center}
\end{table*}

A mesh sensitivity test was conducted for Case 7 which has a moderate rotational speed. The ``Relaxed'' mesh resolution is reduced from ``Standard'' with resolution by a factor of 1.5 in all three directions. Mean heat transfer on the shroud and the two discs for the ``Standard'' and ``Relaxed'' meshes are compared in Table \ref{tab:mesh_sensitivity}. Disc heat transfer is more sensitive to the mesh resolution than the shroud, but the difference is still less than $1.66\%$. Further checks on balance errors, flow structures and quantities show little difference between these two meshes. This indicates the previous mesh guidelines for WMLES could be relaxed for high $\mathrm{Re}_\phi$. The mesh node number is reduced by a factor of $\sim 3$ and the cost is reduced by a factor of $\sim 5$, due to the combination of relaxed mesh and time step. In the following discussions, the $\mathrm{Re}_\phi=2.3\times10^6$ cases are based on the ``Standard'' mesh. But the ``Relaxed'' mesh criteria are used for the highest $\mathrm{Re}_\phi$ case (Case 12, $\mathrm{Re}_\phi=10^7$).

\begin{table}
\caption{Mesh sensitivity test for case 7 (6kR4B24).}
\begin{center}
\label{tab:mesh_sensitivity}
\scalebox{0.9}{
\begin{tabular}{c c c c}
\toprule
Mesh & Shroud $\overline{q}$ & Upstream Disc $\overline{q}$ & Downstream Disc $\overline{q}$\\
Unit & [$\mathrm{W~m^{-2}}$] & [$\mathrm{W~m^{-2}}$] & [$\mathrm{W~m^{-2}}$]\\
\midrule
Standard                & 3841.21 & 491.02 & 541.71 \\
Relaxed                 & 3832.80 & 482.87 & 532.90 \\
Diff.                   & -0.22\% & -1.66\%& -1.63\% \\
\bottomrule
\end{tabular}}
\end{center}
\end{table}

\begin{figure}[!htbp]
	\centerline{\includegraphics[width=\linewidth]{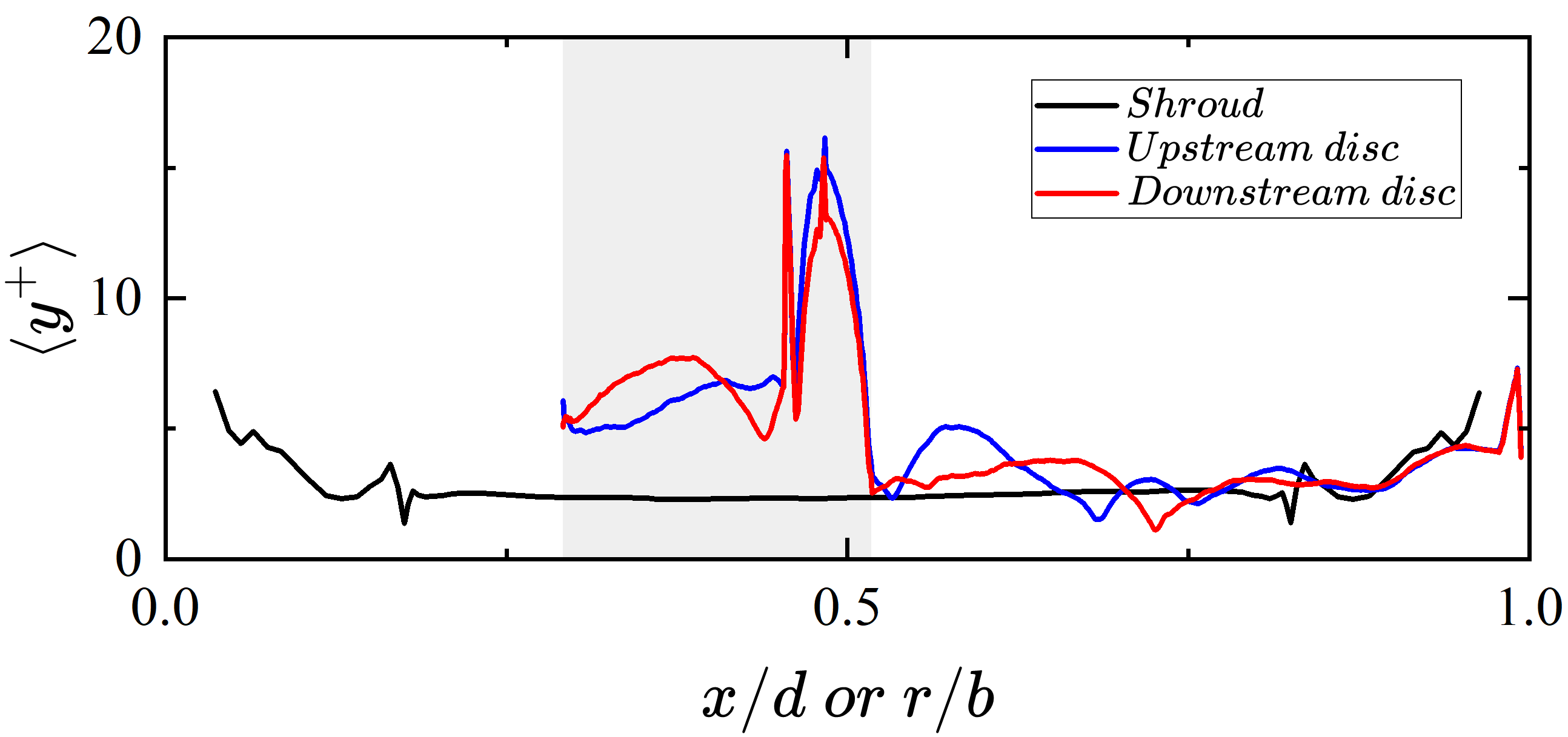}}
	\caption{Mean wall Yplus profiles for Case 12 ($Re_\phi=10^7$) with disc cob and sharp corner regions highlighted in grey.}
	\label{fig:yplus}
\end{figure}

The mean wall Yplus values ($\langle y^+ \rangle$) on shroud and discs for Case 12 ($\mathrm{Re}_\phi=10^7$) are shown in Fig.~\ref{fig:yplus}. Although the near wall grid spacing is relaxed to $1.5\times \delta$, the mean wall Yplus is still reasonably small. Over most of the walls, the mesh extends into the viscous sublayer in which the effect of wall function is largely nullified. On the discs, a region with relatively high and fluctuating mean wall Yplus occurs, as highlighted in grey. This region includes the cobs and the sharp corner between each disc and cob.

\subsection{Convergence}
\label{subsec:convergence}

As listed in Table \ref{tab:mesh}, for all the cases except Case 12 ($\mathrm{Re}_\phi=10^7$), simulations were run for 100 revolutions to reach a statistically steady state, then another 100 revolutions to collect mean statistics. Following checks at lower $\mathrm{Re}_\phi$ conditions, these simulation times were halved for Case 12 ($\mathrm{Re}_\phi=10^7$).

Convergence was examined prior to post-processing. In addition to the widely used criteria, which check equation residuals and variable fluctuations at representative monitor points, further checks were conducted by examining the overall balances of mass flow, angular momentum and energy. Mass flow, angular momentum and energy balance errors are calculated, from the mean values of the last 100 or 50 revolutions after convergence. The balance errors are in percentage of the inlet mass flow rate, angular moment on all the walls and total heat transfer on all the diabatic walls, respectively. The balance errors for representative high speed conditions are listed in Table~\ref{tab:errors}.

Compared to previous studies \cite{wang2023advanced, sun2023parametric, sun2022lowro, gao2022flow} at relatively low speed conditions, the balances indicate very good convergence. This may be associated with the finer mesh and smaller time step used at high Reynolds numbers. Although fewer revolutions were simulated for Case 12 ($\mathrm{Re}_\phi=10^7$), the balance errors are similar. Further checks on the real-time mean shroud heat transfer showed the variation with time to be less than $1\%$ when collecting statistics after 20 revolutions.




\begin{table}
\caption{Balance errors for convergence.}
\begin{center}
\label{tab:errors}
\scalebox{0.85}{
\begin{tabular}{c c c c c}
\toprule
No. & Case Tag & Mass Flow & Angular Momentum & Energy\\
\midrule
7 & 6kR4B24    & 0.03\%  & 2.29\%          & -1.25\%  \\
11& 8kR4B24    & 0.01\%  & 0.32\%          &  -1.44\% \\
12&Re1E7R4B24 & 0.00\%  & -3.20\%         & 1.90\%   \\
\bottomrule
\end{tabular}}
\end{center}
\end{table}

\subsection{Computational Cost and Parallel Performance}
\label{subsec:cost}
\begin{figure}[!htbp]
	\centerline{\includegraphics[width=\linewidth]{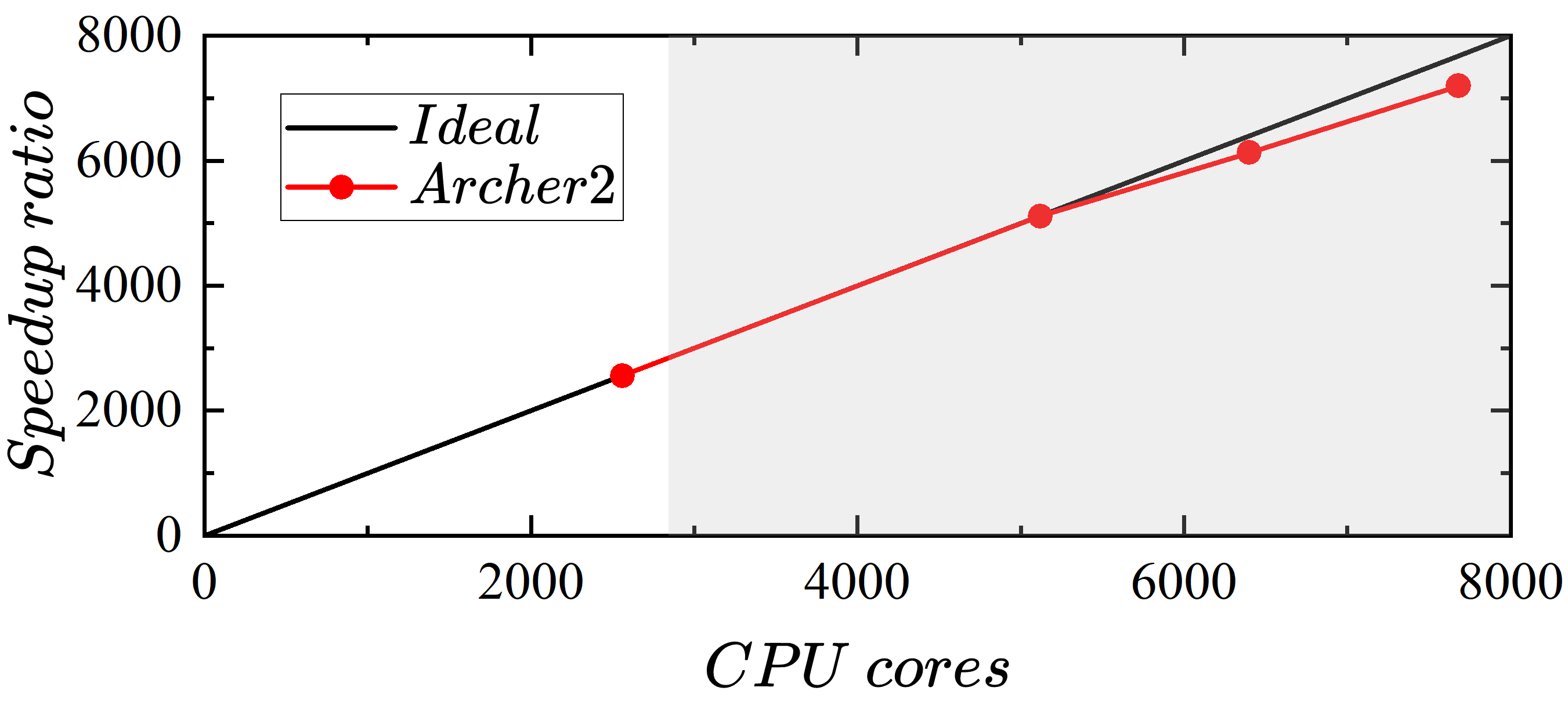}}
	\caption{Parallel performance on the national HPC facility for Case 12 ($Re_\phi=10^7$) with $\sim 2.27\times10^8$ mesh nodes.}
	\label{fig:parallel}
\end{figure}

In Fig.~\ref{fig:parallel}, parallel performance on the national HPC facility ARCHER2 is shown, for~Case 12 ($\mathrm{Re}_\phi=10^7$) with $\sim 2.27\times10^8$ mesh nodes. The recommended distribution of 20k to 80k mesh nodes per CPU core \cite{edf2022guide} is highlighted as a grey zone. Starting with 2560 CPU cores (AMD EPYC$^\text{TM}$ 7742, 2.25 GHz) as a baseline, increasing to 5120, 6400 and 7680 CPU cores give speedup ratios $99.8\%$, $95.8\%$ and $93.7\%$, respectively. This confirms excellent parallel performance and shows improvements compared to the local HPC cluster \cite{wang2023advanced}.

\begin{figure}[!htbp]
	\centerline{\includegraphics[width=\linewidth]{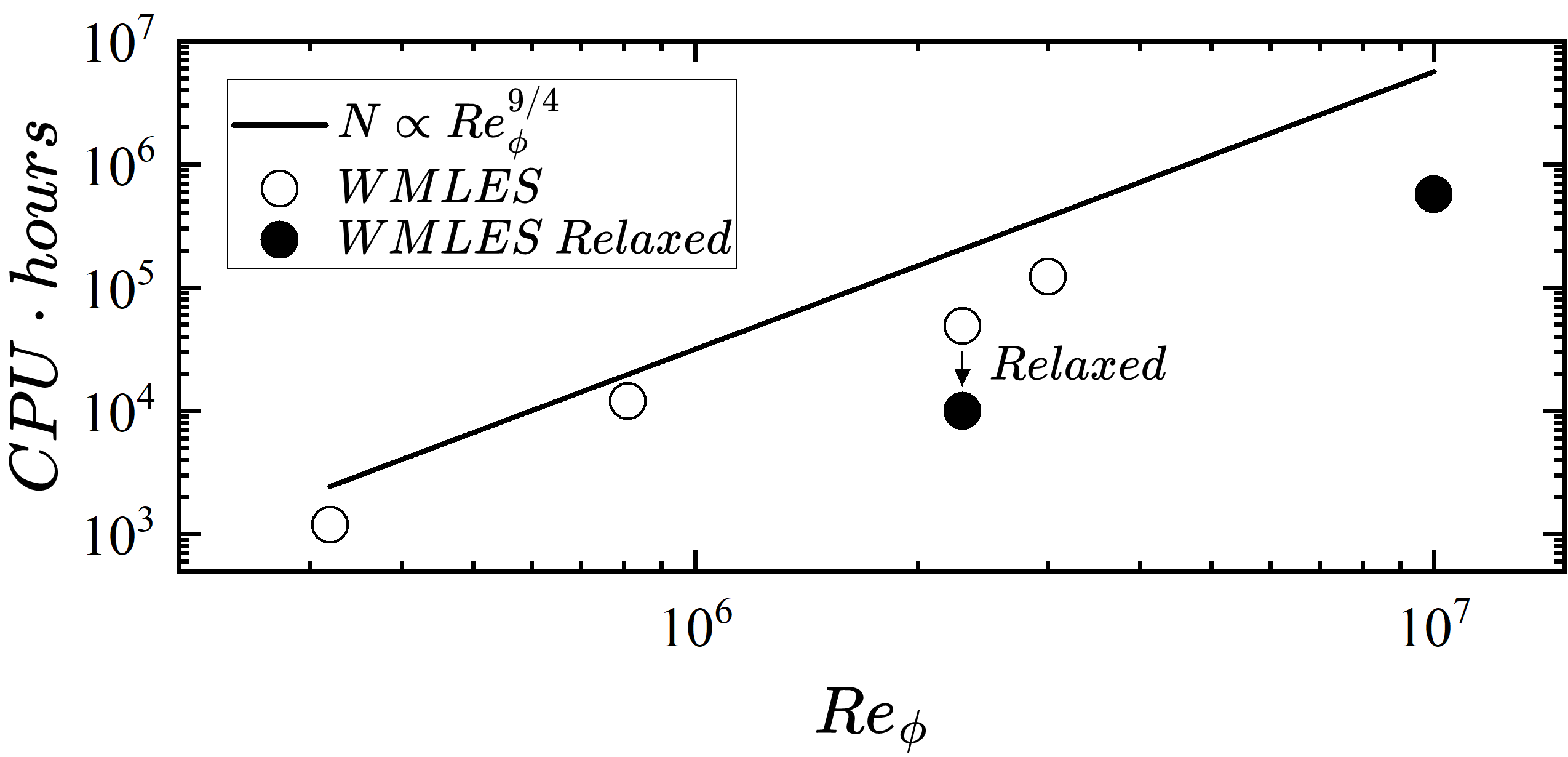}}
	\caption{Scaling of computational cost with Reynolds number.}
	\label{fig:cost}
\end{figure}

The computational cost in CPU hours for WMLES varies with Reynolds number. These are listed in Table~\ref{tab:mesh} and shown in Fig.~\ref{fig:cost}. The scaling $N \propto \mathrm{Re}_\phi ^{9/4}$ is an estimate of the increase of mesh size with Reynolds number for direct numerical simulation (DNS), and is included for reference. Considering the necessity of a smaller time step for a finer mesh, the cost of DNS would increase at a greater rate. The costs achieved confirm the potential for exploiting WMLES in engineering applications, and Fig.~\ref{fig:cost} may be used to estimate the computational cost. The turnaround time for the $\mathrm{Re}_\phi=10^7$ case was $\sim$ 74 hours ($\sim$ 3 days) using 7680 CPU cores.

\section{Results and Discussions}
\label{sec:results}
\subsection{Comparison with Previous Data}
\label{subsec:validation}

Mean shroud heat transfer is compared with experimental data \cite{jackson2021analysis, jackson2021measurement}, and previous WMLES data \cite{gao2022flow} where available, in Table~\ref{tab:shroud_Nu_validation}.~In the experiments, shroud heat transfer is measured by two RdF 27160-C-L-A01 thermopile heat flux gauges, each individually calibrated for steady-state measurements between 0.5 and 8$\mathrm{kW/m^2}$ and gauge temperatures up to 110$^\circ$C \cite{jackson2021measurement}. The Nusselt number here ($\mathrm{Nu}_b^\prime$) is defined using the rotary stagnation temperature difference between shroud and inlet. At low speed conditions (Cases 1 and 2), previous work has suggested that WMLES may underpredict turbulence levels in some areas \cite{wang2023advanced, gao2022flow}. This may account for some differences with measurements and previous WMLES in these cases. As previously discussed for Case 1 \cite{wang2023advanced}, the present WMLES are closer to experiments than results from a full compressible solver \cite{gao2022flow}.

At moderate speed, with Rossby number 0.2 (Cases 3 and 4), the agreement with experiment is very good, with differences below 3\%. The unsteady RANS (Case 5, highlighted in red), which used the same mesh and time step as the WMLES, gave larger differences with the experiment of around 10\%. This is consistent with the advantage of WMLES over URANS found in previous studies \cite{wang2023advanced, sun2006les, owen2007buoyancy}.

At moderate speed, with Rossby number 0.4 (Cases 6 and 7), WMLES gives shroud heat transfers significantly lower than the experiments. However, results for the present and previous WMLES for Case 7 are close. Jackson et al. \cite{jackson2022unsteady} measured peaks in cavity unsteady pressure amplitude at $\mathrm{Ro}=0.5$ with amplitudes reducing for $\mathrm{Ro}<0.4$. Previous compressible flow WMLES diverged due to flow reversal at the inlet and outlet for $\mathrm{Ro}<0.4$ using the present solution domain. Extending and contracting the inlet and outlet boundaries allowed solutions to be obtained \cite{sun2022lowro}.~The inlet and outlet conditions used in the present solver appear less prone to divergence but may limit flow unsteadiness.

Case 9, highlighted in blue, extends the validation to the highest speed of 8000 rpm considered in the experiments, and selects the highest flow rate at this speed, corresponding to Ro = 0.2. The WMLES shows very good agreement with the experiment, with a difference of 1.2\%. For this operating condition, unsteady RANS (Case 10, highlighted in red) shows a strong under-prediction of $\sim 18\%$.

\begin{table}[!htbp]
	\caption{Comparison of shroud mean Nusslet number with reference data.}
	\begin{center}
		\label{tab:shroud_Nu_validation}
		\scalebox{0.8}{
   \begin{tabular}{c c c |c c| c c c}
				\toprule
				No. & Case Tag & Exp. & WMLES\cite{gao2022flow} & Diff. & Present & Diff. \\
				\midrule
	          1&0.8kR5B26 & 11.61 & 8.82 & -24.0\%& 12.63  & 8.8\% \\			
                2&2kR4B26   & 25.98 & 22.32& -14.1\%& 28.20  & 8.6\%  \\
				3&6kR2B18   & 36.49 & [-]  & [-]     & 37.07  & 1.6\%  \\
				4&6kR2B26   & 48.66 & [-]  & [-]     & 50.09  & 3.0\%  \\
				\textcolor{red}{5}&\textcolor{red}{6kR2B26-U} & \textcolor{red}{48.66} & \textcolor{red}{[-]}  & \textcolor{red}{[-]}     & \textcolor{red}{43.92}  & \textcolor{red}{-9.7\%} \\
                6&6kR4B18   & 50.95 & [-]  & [-]     & 39.99  & -21.5\%\\
				7&6kR4B24   & 55.06 & 47.85& -13.10\%& 46.53  & -15.5\%\\
				\textcolor{blue}{9}&\textcolor{blue}{8kR2B24} & \textcolor{blue}{53.13} & \textcolor{blue}{[-]}  & \textcolor{blue}{[-]}     & \textcolor{blue}{52.48}  & \textcolor{blue}{-1.2\%}  \\
                \textcolor{red}{10}&\textcolor{red}{8kR2B24-U} & \textcolor{red}{53.13} & \textcolor{red}{[-]}  & \textcolor{red}{[-]} & \textcolor{red}{43.61}  &  \textcolor{red}{-17.9\%}\\
				\bottomrule
		\end{tabular}}
	\end{center}
\end{table}

\begin{figure}[!htbp]
	\centerline{\includegraphics[width=.975\linewidth]{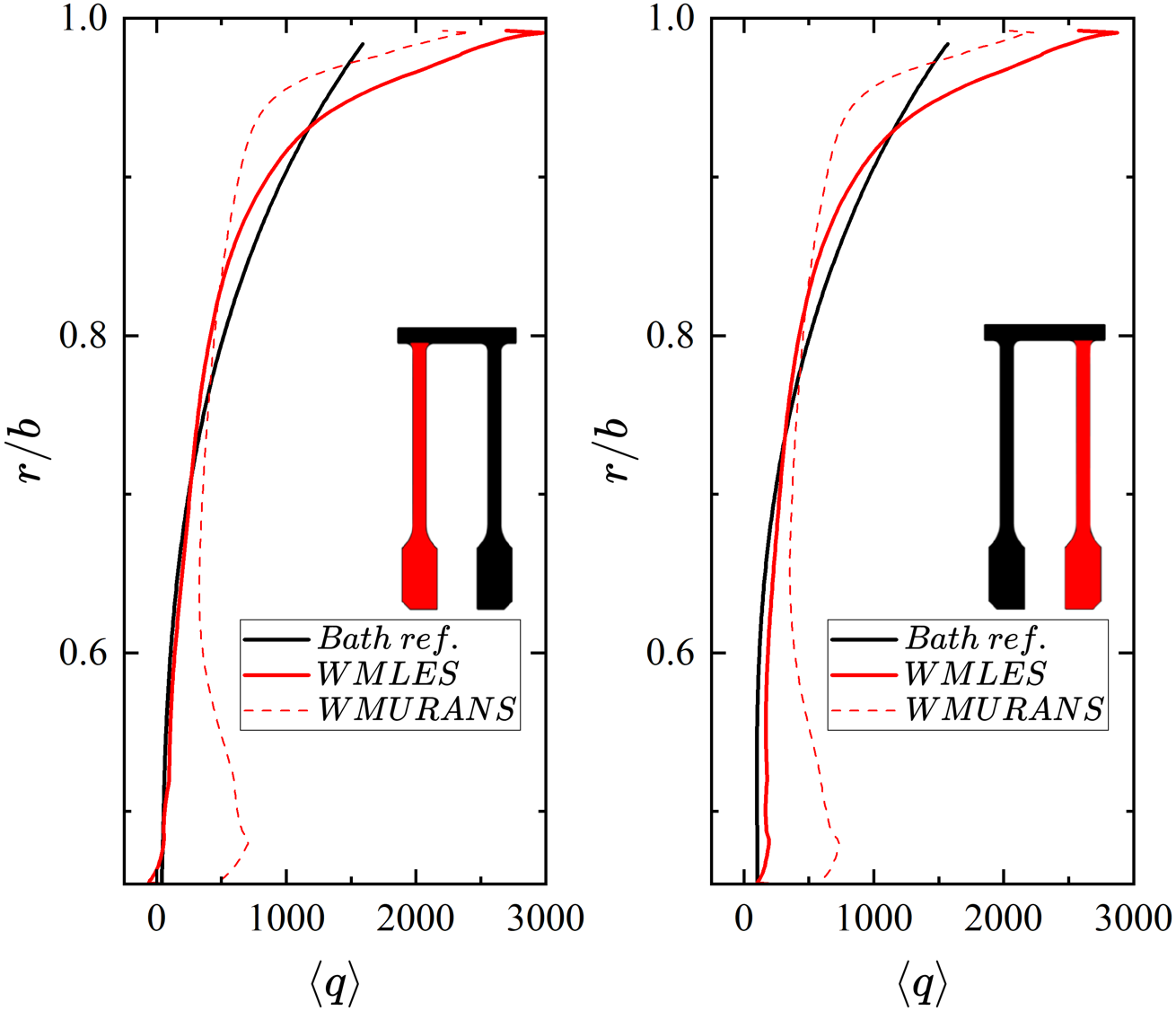}}
	\caption{Mean convective heat flux [$\mathrm{W~m^{-2}}$] on discs for Cases 9 (WMLES) and 10 (WMURANS). The left sub-figure shows the upstreamstream disc; the right sub-figure shows the downstream disc.}
	\label{fig:disc_q_validation}
\end{figure}

Mean convective heat fluxes on upstream and downstream discs are compared with the experimental data in Fig.~\ref{fig:disc_q_validation}, for experimental cases at the highest rotational speed and buoyancy parameter. In the experiments \cite{jackson2021measurement}, the disc temperatures were measured from a radial distribution of 28 K-type thermocouples on both upstream and downstream discs.~At high radii ($r/b>0.6$), the temperatures on the upstream and downstream discs mainly agreed within the experimental uncertainties.~Time-averaged disc temperatures were used in a Bayesian model, to provide an experimentally derived continuous radial variation. The Bayesian model, in conjunction with the circular-fin equation model, was used to determine the variation of heat flux with radius. The radiative heat flux was subtracted from the total heat flux to get the convective heat flux, as shown with the black curves in Fig.~\ref{fig:disc_q_validation}. In the experimental reference \cite{jackson2021measurement}, the 95\% confidence interval indicates the uncertainty of the convective heat flux within $\sim \pm 4\%$ of the maximum convective heat flux for $r/b$ between 0.5 and 0.9. The uncertainty rises near the disc cob and shroud.

WMLES and WMURANS data are averaged in time and in the circumferential direction. WMLES and experimental data are in good agreement over the discs for $r/b<0.9$. Higher disc heat flux is shown than the experimental data at high radii close to the shroud. This was also noted in previous WMLES \cite{wang2023advanced, gao2022flow}, and could be related to uncertainty in deducing heat flux from the disc radial temperature gradients in this region. WMURANS gives significantly larger differences with the experimental data, which extends the previous evidence under low Reynolds number conditions.



\begin{figure}[!htbp]
	\centerline{\includegraphics[width=.975\linewidth]{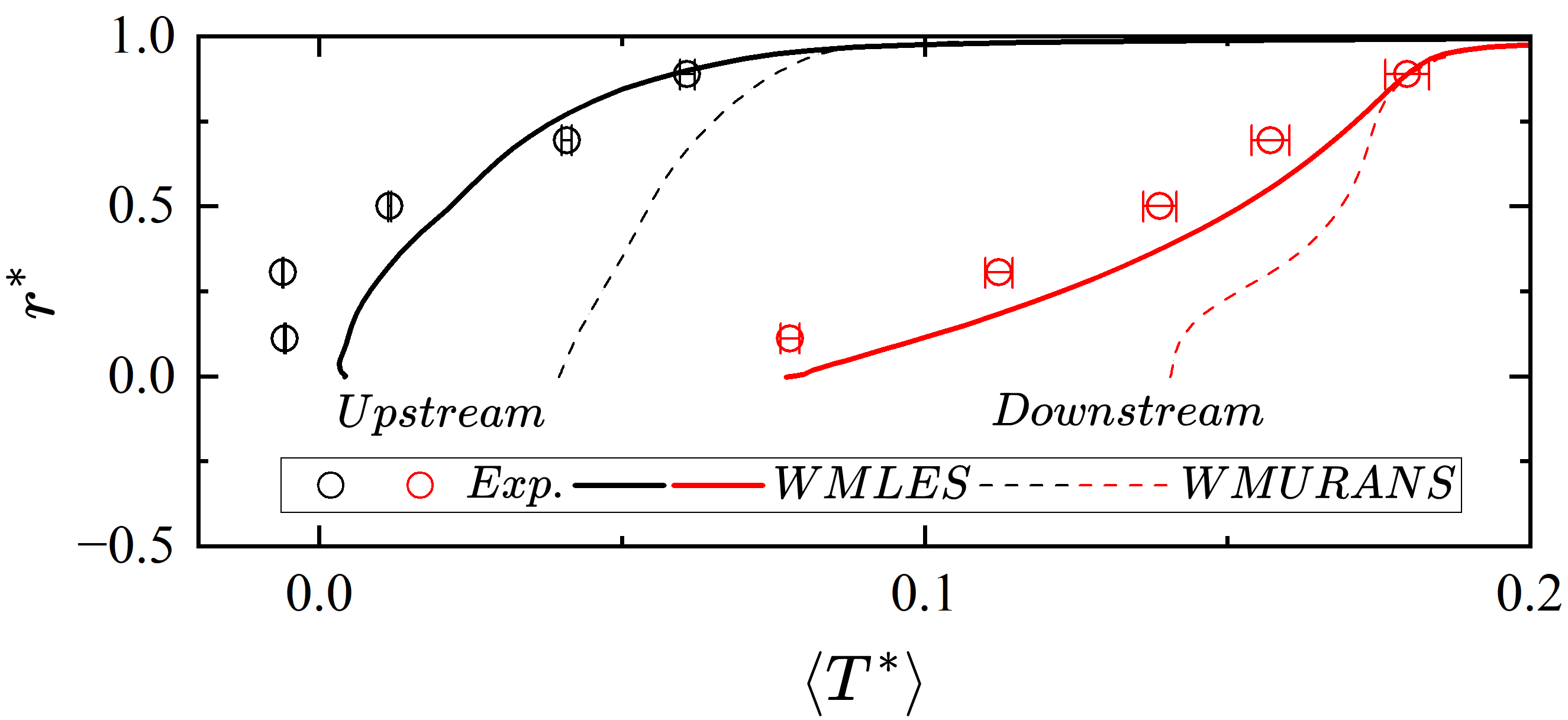}}
	\caption{Bore flow temperature for Cases 9 (WMLES) and 10 (WMURANS).}
	\label{fig:bore_flow_T}
\end{figure}

WMLES and WMURANS data are compared with measured bore flow temperatures for the same condition in Fig.~\ref{fig:bore_flow_T}. Temperature profiles are shown radially from the shaft to the upstream and downstream disc bores. The error bars indicate uncertainty of $\pm$ 1K in flow temperature measurements. WMLES shows very encouraging agreement with measurements, considering the possibility of sensitivity to boundary conditions. Again, WMURANS shows significant departures.


\begin{figure}[!htbp]
	\centerline{\includegraphics[width=.975\linewidth]{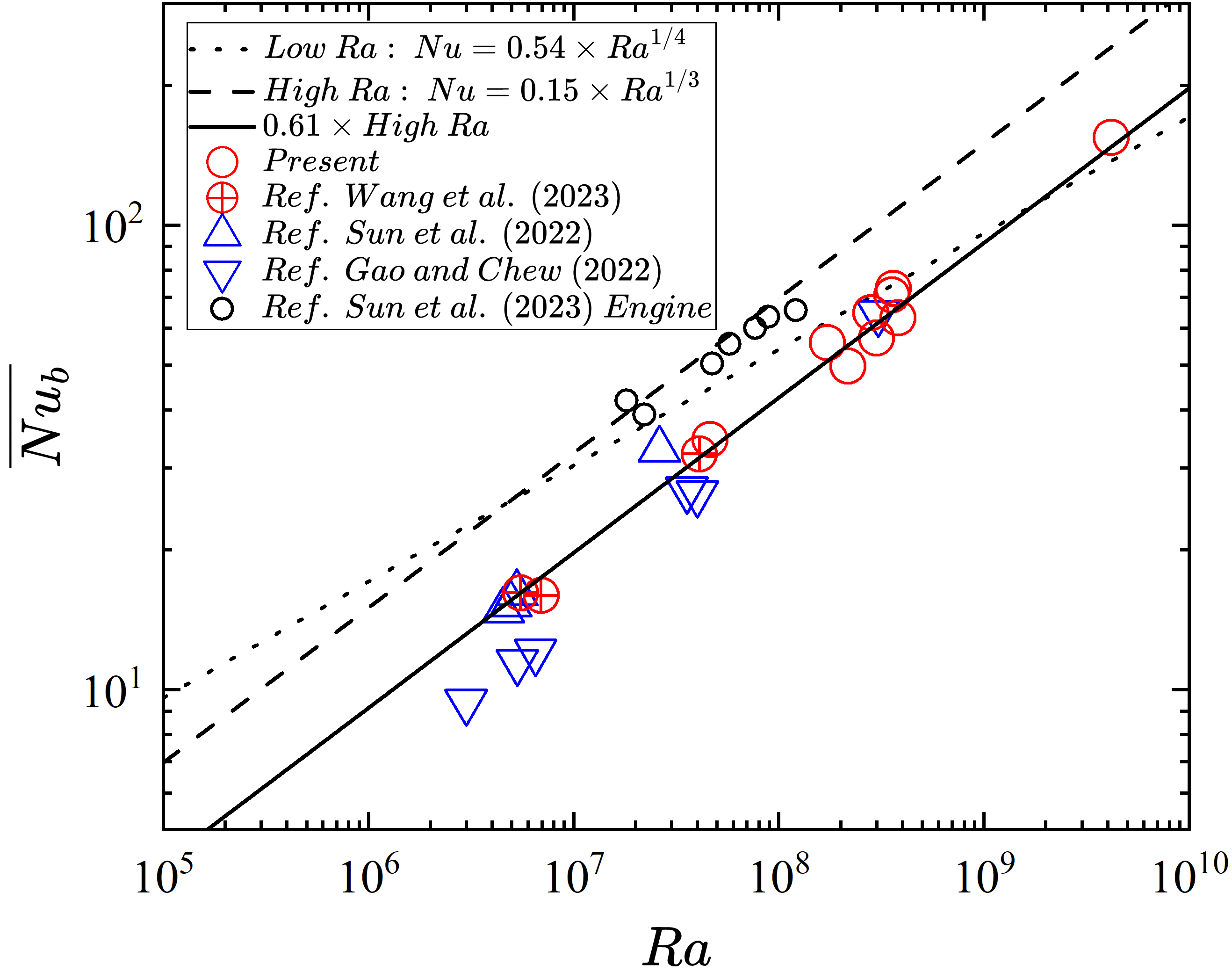}}
	\caption{Shroud mean Nusselt number for all WMLES cases, comparing with recently published numerical data and experimental correlations.}
	\label{fig:shroud_Nu_scale}
\end{figure}


\subsection{Mean Shroud and Disc Heat Transfer}


Shroud mean Nusselt numbers ($\mathrm{Nu}_b$) are plotted in Fig.~\ref{fig:shroud_Nu_scale}. Here the rotary stagnation temperature difference between the shroud and core is used in defining $\mathrm{Nu}_b$ and Ra. The core temperature in the present results is the average of mean $T_r$ over $x/d=0.25$ to 0.75 at $r/b=0.9$. For comparison, previous WMLES for the same test rig \cite{gao2022flow, sun2022lowro, wang2023advanced} and WMLES for an engine disc cavity geometry \cite{sun2023parametric} are also plotted. The dotted and dashed lines are adaptations of established low and high Rayleigh number experimental correlations for gravitational convection. The solid line is 0.61 times the value for the high Rayleigh number correlation. This was found to produce a reasonable fit of experimental data to an elementary model as described in Section~\ref{sec:intro} Fig.~\ref{fig:Xshroud_Xdisc}. Note that core temperatures were not measured in the experiments, so direct comparisons of $\mathrm{Nu}_b$ with experiments are not possible.


The results for the present test rig show scaling of $\mathrm{Nu}_b$ with Ra close to that for the high Ra correlation ($\mathrm{Nu \propto Ra ^ {1/3}}$), and the values are consistent with the interpretation of the experimental measurements ($\mathrm{Nu = 0.61 \times 0.15 \times Ra ^ {1/3}}$), given in Fig.~\ref{fig:Xshroud_Xdisc}. At relatively low Ra, the present data is higher than earlier results from Gao and Chew \cite{gao2022flow}, but agrees well with those from Sun et~al. \cite{sun2022lowro}. Data for the engine geometry and operating conditions are consistently higher than those for the test rig, and closer to the standard high Ra correlation. Differences between the engine and test rig may come from one or a combination of differences in cavity geometry, inlet swirl, Rossby number and surface temperature distributions.

\begin{figure}[!htbp]
	\centerline{\includegraphics[width=\linewidth]{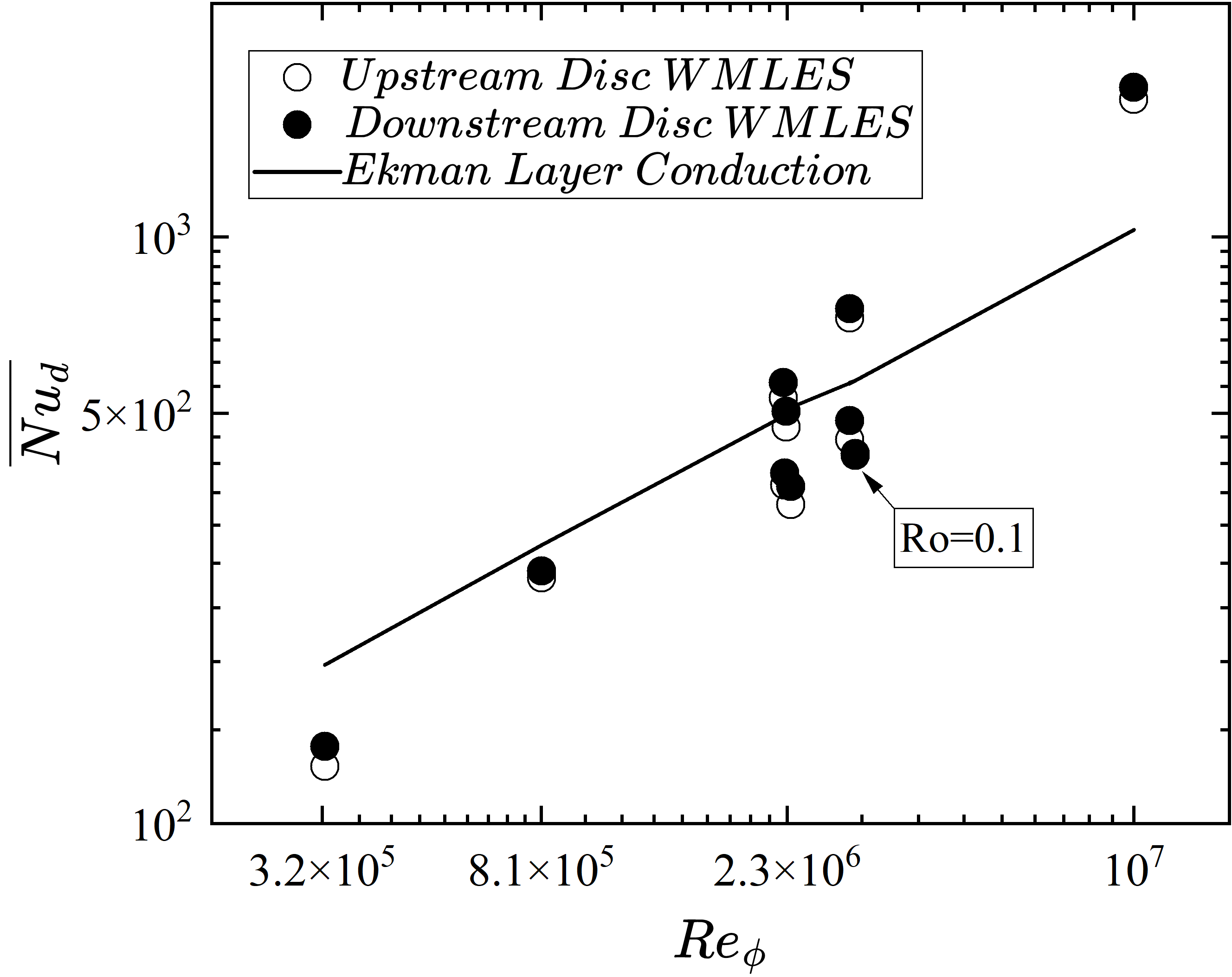}}
	\caption{Disc mean Nusselt number ($Nu_d = \langle q_d \rangle b /(\lambda_d (T_{r,d} - \langle T_{r,c} \rangle))$) for all WMLES cases.}
	\label{fig:disc_mean_q_with_EL}
\end{figure}

Disc mean Nusselt numbers ($Nu_d$), defined using the rotary stagnation temperature difference between disc mean and core, are plotted in Fig.~\ref{fig:disc_mean_q_with_EL}.~Values assuming Ekman layer conduction are calculated by substituting $\langle q_d \rangle$ by $\langle q_{\mathrm{EL}} \rangle = \langle \lambda_d (T_{r,d}-\langle T_{r,c} \rangle) \rangle / \Delta$ in the definition of $Nu_d$, giving the curve with $\mathrm{Nu} \propto \mathrm{Re}_\phi ^ {1/2}$. For all cases, the mean heat transfer on the downstream disc is slightly higher than that on the upstream disc. This is mainly related to the heat transfer enhancement by impingement on the downstream disc cob. For the lowest Rossby number case (Case 8, Ro=0.1), in which the impingement is expected to be the weakest, mean heat transfer on the upstream disc has the smallest difference from that on the downstream disc.


Previous work has shown correspondence of measurements and WMLES disc heat transfer with the estimate from Ekman layer conduction. However, the WMLES may under-predict turbulence and disc heat transfer at low Reynolds number conditions \cite{gao2022flow}. This may explain the relatively low $\mathrm{Nu}_d$ for $\mathrm{Re}_\phi=3.2\times10^5$ in Fig.~\ref{fig:disc_mean_q_with_EL}. Results for $\mathrm{Re}_\phi=8.1\times10^5 - 3.0\times10^6$ show better agreement with Ekman layer conduction, although there is significant variation with operating conditions. This, and the shroud mean heat transfer in Fig.~\ref{fig:shroud_Nu_scale}, are consistent with the scatter found in the analysis of the experimental results in Fig.~\ref{fig:Xshroud_Xdisc}.~At the highest Reynolds number ($10^7$), the WMLES clearly gives higher disc heat transfer than the Ekman layer conduction estimate. This operating condition, which is well beyond the experimental test matrix, is further examined below.

\subsection{Reynolds Number Effects}
\label{subsec:highRe}
 
\begin{figure}[!htbp]
\centerline{\includegraphics[width=.9\linewidth]{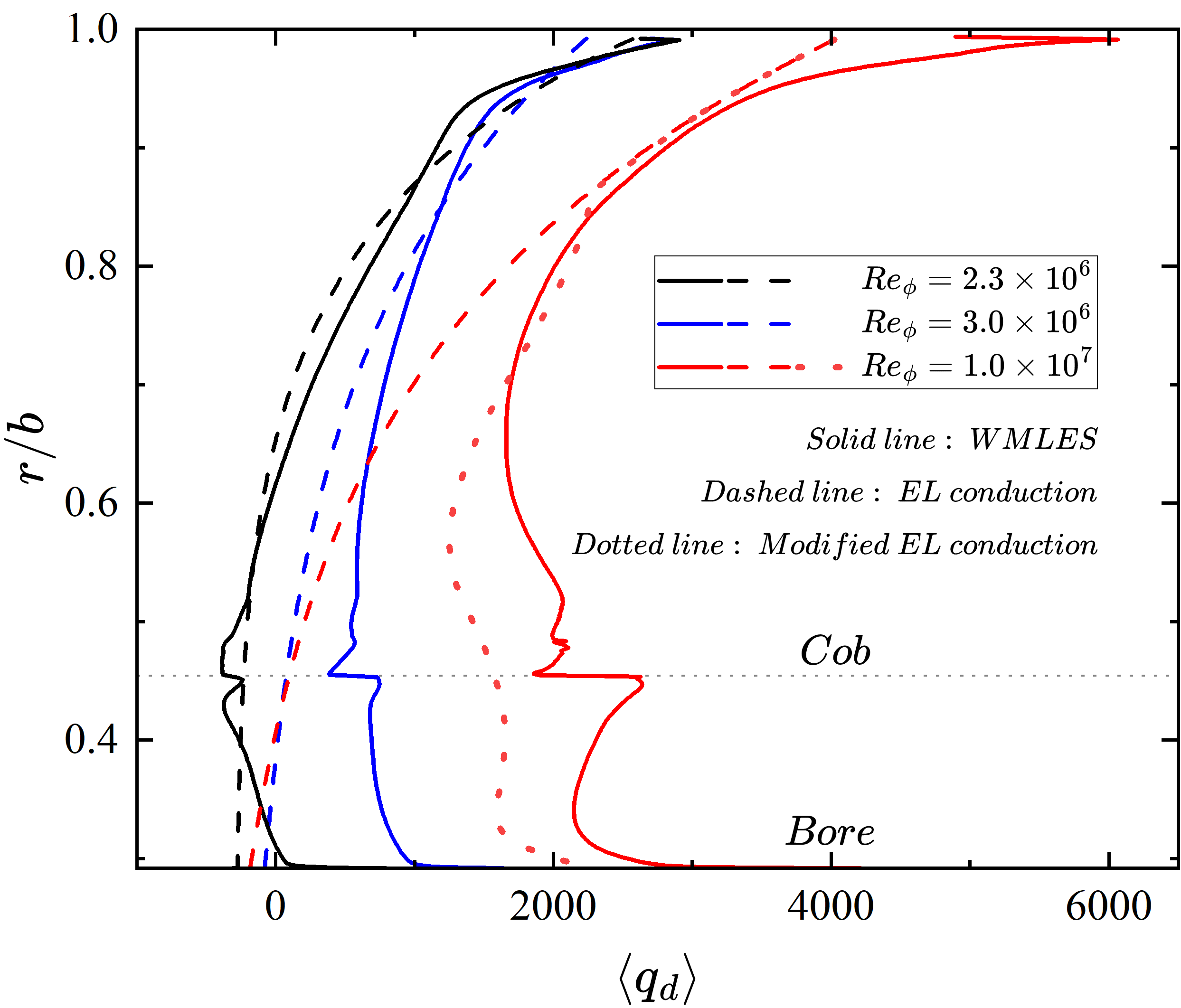}}
	\caption{Mean convective heat flux [$\mathrm{W~m^{-2}}$] on the upstream disc for Cases 7, 11 and 12 ($\mathrm{Ro=0.4}$).}
	\label{fig:disc_local_q_with_EL}
\end{figure}

Figure \ref{fig:disc_local_q_with_EL} shows the mean heat transfer distribution on the upstream disc for Cases 7, 11 and 12. Profiles on the downstream disc show similar trends. WMLES results are compared with estimates from Ekman layer conduction using mean core temperatures from the simulations. A progressive departure from the Ekman layer conduction estimate is seen as $\mathrm{Re}_\phi$ increases. As expected from Fig.~\ref{fig:disc_mean_q_with_EL}, for $\mathrm{Re}_\phi=10^7$, the WMLES disc heat transfer is well above the estimate. The departure is greatest at lower radii on the disc. Note that for matching experimental conditions, the two higher $\mathrm{Re}_\phi$ cases have a different disc temperature distribution from the $\mathrm{Re}_\phi=2.3\times10^6$ case. This could contribute to the observed trends, but is not considered to be the primary cause.

The modified Ekman layer conduction curve for $\mathrm{Re}_\phi=10^7$ in Fig.~\ref{fig:disc_local_q_with_EL} applies a correction from turbulent Ekman layer theory \cite{Owen_Pincombe_Rogers_1985}. The laminar value is calculated using a local core temperature and multiplied by the ratio of circumferential wall shear in turbulent and laminar Ekman layers. For the boundary layer Reynolds numbers ($\mathrm{Re_\delta}$) above 180, the multiplying factor is $(\mathrm{Re_\delta} / 180)^{0.6}$. The modification brings the theoretical prediction much closer to the WMLES at lower radii. Further discussion of transition and calculation of $\mathrm{Re_\delta}$ is given below.

\begin{figure}[!htbp]
     \centering
     \begin{subfigure}[b]{0.4\textwidth}
         \centering
         \includegraphics[width=\textwidth]{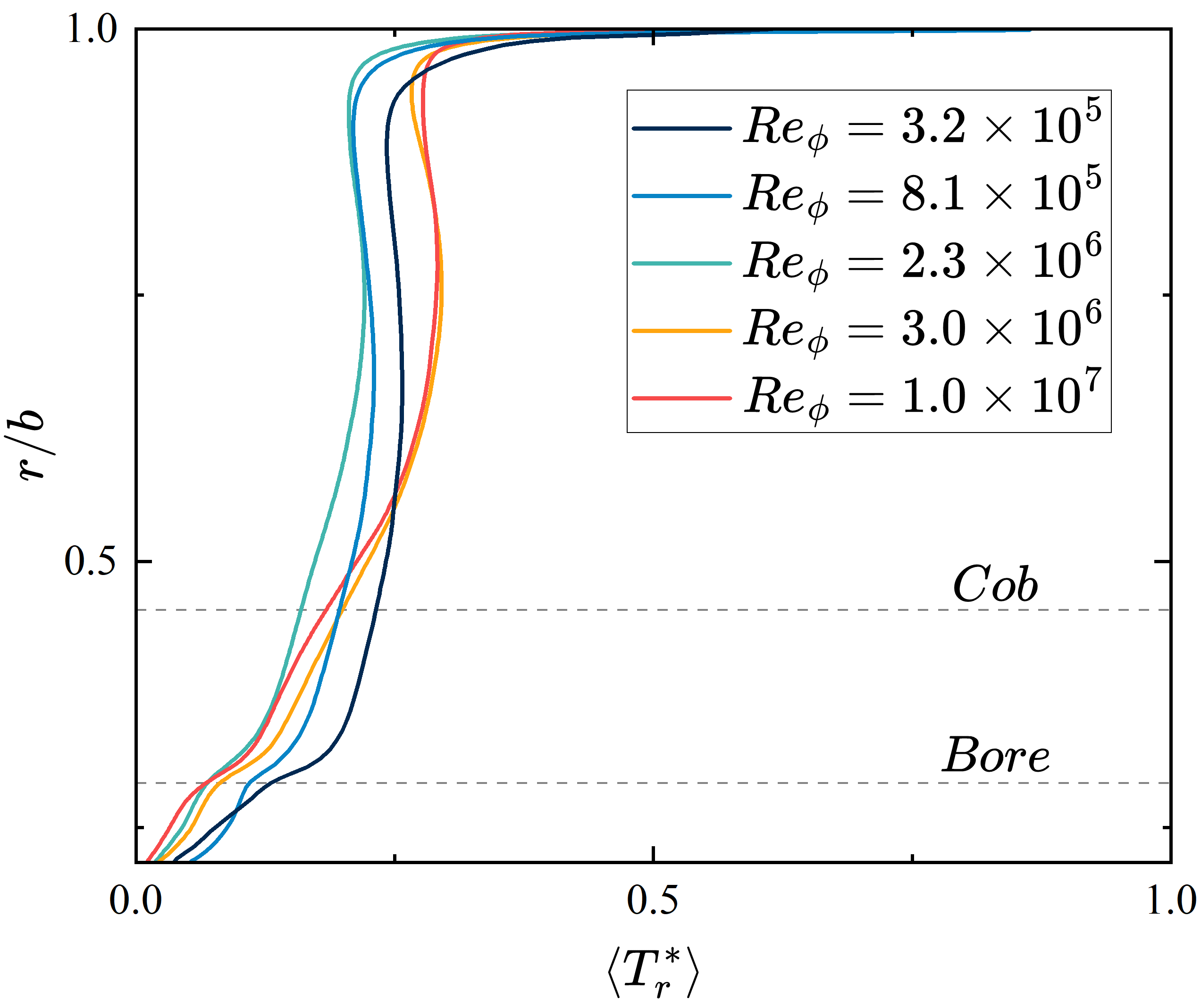}
     \end{subfigure}
     \hfill
     \begin{subfigure}[b]{0.4\textwidth}
         \centering
         \includegraphics[width=\textwidth]{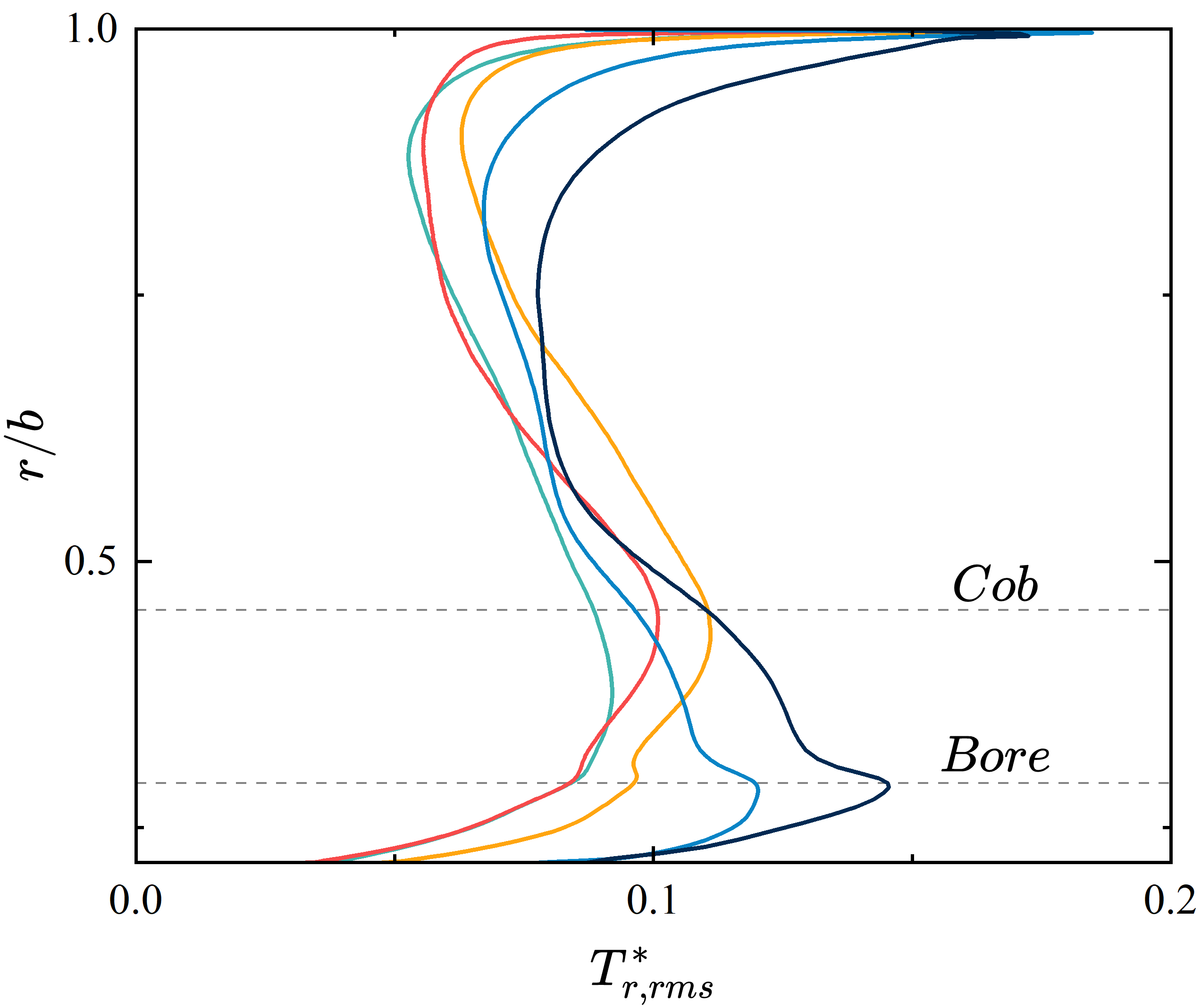}
     \end{subfigure}
        \caption{Profiles of mean rotary stagnation temperature and its fluctuation at mid-axial position for Cases 1, 2, 7, 11 and 12.}
        \label{fig:mid_axial_Tr_Trrms}
\end{figure}

 
There is also evidence of changes in the cavity mean rotary stagnation temperature ($\langle T_r \rangle$) distribution at higher $\mathrm{Re}_\phi$. Figure~\ref{fig:mid_axial_Tr_Trrms} shows mean rotary stagnation temperature and its fluctuation at cavity mid-axial position, covering the full range of $\mathrm{Re}_\phi$. The values are first time-averaged at the local position, then circumferential-averaged. The $\langle T_r^* \rangle$ profiles for the two higher $\mathrm{Re}_\phi$ cases show a stronger variation from $r/b$ around 0.3 to 0.6 than the other three cases. The form of the fluctuation distribution in the disc cob region is also different for these two higher $\mathrm{Re}_\phi$ cases. The reduced uniformity in core mean rotary stagnation temperature may be associated with stronger disc cooling at lower radii and subtle changes in the unsteady flows.

\begin{figure}[!htbp]
     \centering
     \begin{subfigure}[b]{0.375\textwidth}
         \centering
         \includegraphics[width=\textwidth]{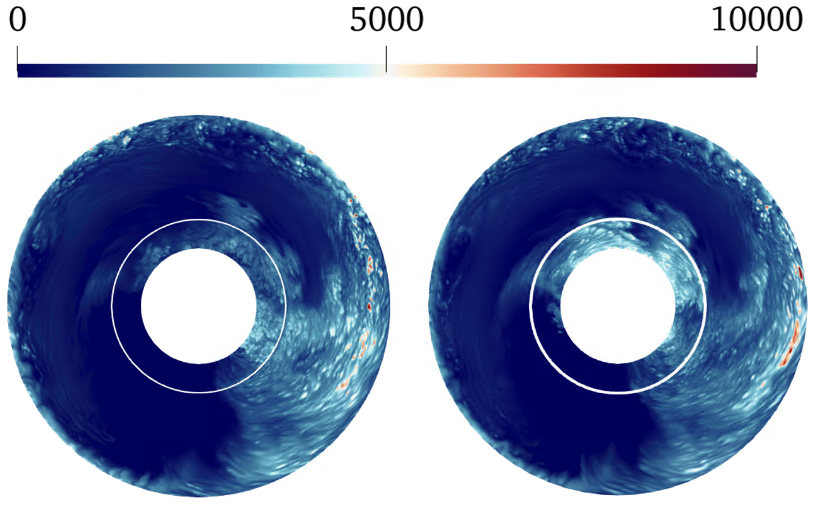}
         \caption{$\mathrm{Re}_\phi=3.0\times10^6$ (Case 11).}
     \end{subfigure}
     \hfill
     \begin{subfigure}[b]{0.375\textwidth}
         \centering
         \includegraphics[width=\textwidth]{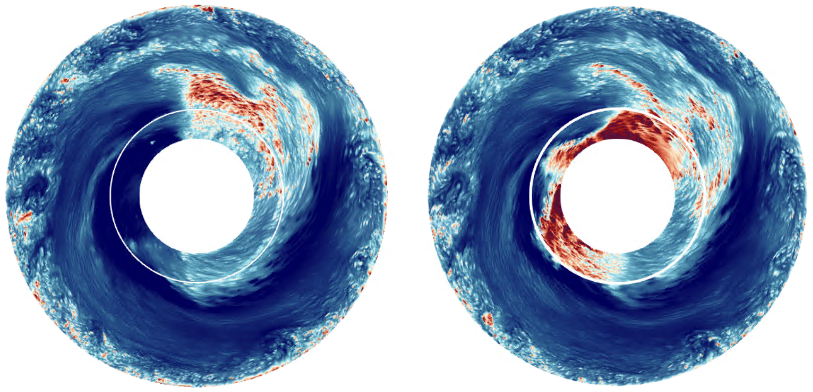}
         \caption{$\mathrm{Re}_\phi=10^7$ (Case 12).}
     \end{subfigure}
        \caption{Instantaneous wall heat flux [$\mathrm{W~m^{-2}}$] on upstream disc (left) and downstream disc (right).}
        \label{fig:instan_q_contours}
\end{figure}

\begin{figure}[!htbp]
	\centerline{\includegraphics[width=.8\linewidth]{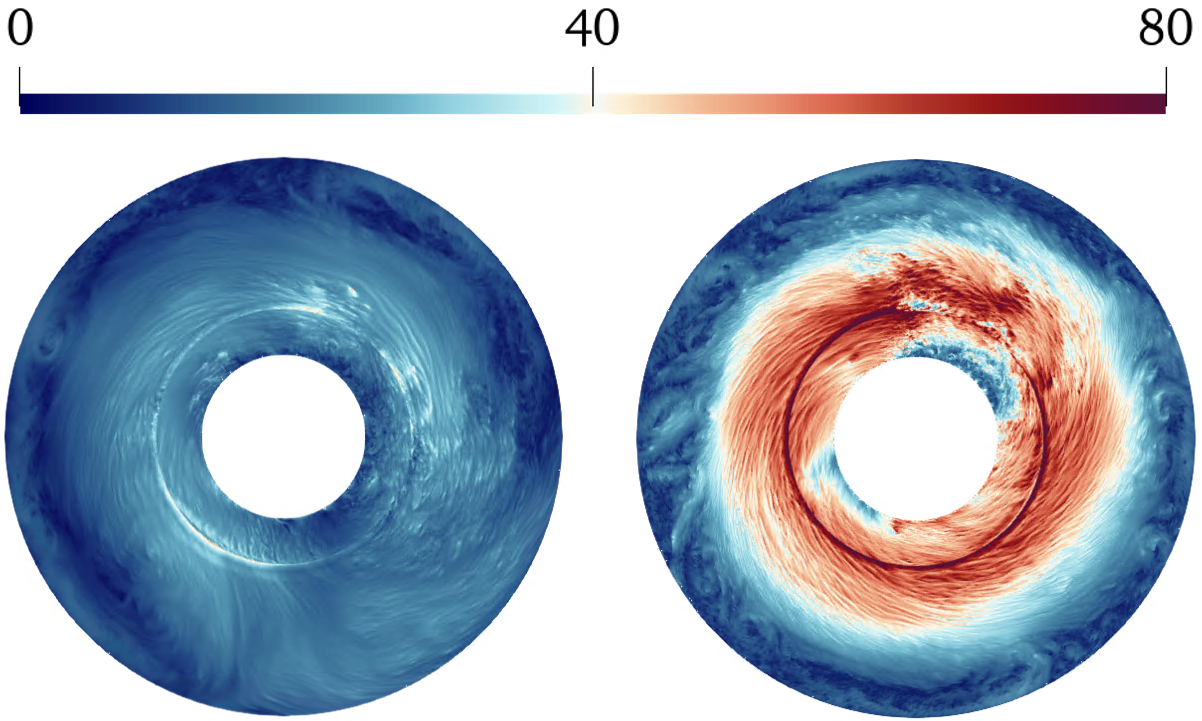}}
	\caption{Instantaneous wall Yplus on upstream disc for $\mathrm{Re}_\phi=3.0\times10^6$ (Case 11, left) and $\mathrm{Re}_\phi=10^7$ (Case 12, right).}
	\label{fig:disc_yplus}
\end{figure}

Further details for the two higher $\mathrm{Re}_\phi$ cases are given by the instantaneous disc heat flux in Fig.~\ref{fig:instan_q_contours} and the instantaneous disc wall Yplus in Fig.~\ref{fig:disc_yplus}. Wall Yplus shows similar trends between upstream and downstream discs, so only those on the upstream disc are shown. The heat flux contours reflect the expected flow structure, with high heat transfer indicating the path of a cold radial outflow plume. For $\mathrm{Re}_\phi=10^7$, the wall Yplus values at lower radii are considerably higher than those for the laminar Ekman layer flows at lower $\mathrm{Re}_\phi$. Spiral structures are apparent in the wall Yplus contours. Similar structures have been observed in WMLES for a rotating cavity with radial outflow, as was used as a test case for heat transfer prediction in validating the present computational code \cite{wang2023advanced}.

\begin{figure}[!htbp]
	\centerline{\includegraphics[width=.8\linewidth]{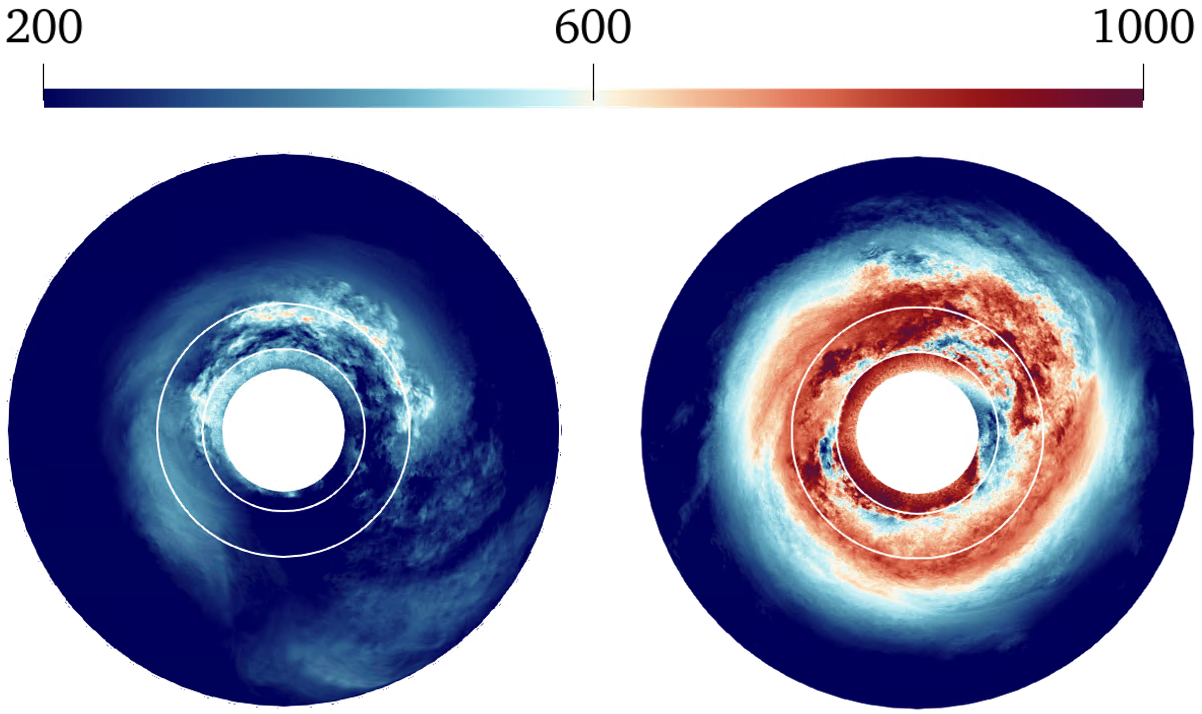}}
	\caption{Instantaneous boundary layer Reynolds number ($\mathrm{Re}_\delta$) at mid-axial position for $\mathrm{Re}_\phi=3.0\times10^6$ (Case 11, left) and $\mathrm{Re}_\phi=10^7$ (Case 12, right).}
	\label{fig:Re_local}
\end{figure}

Laminar-turbulent transition in rotating disc boundary layers has been considered analytically by Lingwood \cite{lingwood_1997}. For a core flow rotating as a forced vortex at a speed below that of the disc, stabilty theory predicts transition above critical values of the boundary layer Reynolds number. These critical values range from 198 for linear Ekman layers to 507 for a free disc. For comparison with these values, contours of instantaneous boundary layer Reynolds numbers ($\mathrm{Re}_\delta$) based on Ekman depth and the relative swirl velocity at the cavity mid-axial position are shown in Fig.~\ref{fig:Re_local}. The values confirm that turbulent disc boundary layers are expected at lower radii. The higher boundary layer Reynolds numbers in these regions are associated with slip of the core flow relative to the rotating disc. In the present cases, no inlet swirl is imposed in the throughflow, which is consistent with the experimental condition but may be different at engine conditions. With inlet swirl, the transition to turbulence is likely to be delayed.

\section{Conclusion}
\label{sec:conclusions}
This research focuses on flow and heat transfer in a compressor rotating disc cavity with axial throughflow, with high rotational speeds (up to 8000 rpm) and high Reynolds numbers (up to $\mathrm{Re}_\phi=10^7$). The main conclusions and contributions are summarised as follows.

\begin{itemize}
  \item [1)]
Validation of wall-modelled large-eddy simulations (WMLES) for compressor disc heat transfer modelling has been extended to the highest Reynolds number for which experimental data is available ($\mathrm{Re}_\phi=3.0\times10^6$). The validity of the Boussinesq approximation and the kinetic energy treatment with rotary stagnation temperature are also confirmed, although these simplifications were adopted here for convenience and are not essential.

  \item [2)]
Based on mesh sensitivity tests, modified meshing criteria for high Reynolds number simulations have been proposed. Combined with the excellent parallel performance of the open-source code on the national high-performance computing facility, this has allowed a simulation at $\mathrm{Re}_\phi=10^7$ to be conducted with a turnaround time of about 3 days with 7680 CPU cores.

  \item [3)]
The present results show consistency with other recent studies and confirm the general flow structures and expected scaling of shroud heat transfer. In the present study, at higher Reynolds numbers, the extent of the isothermal core is reduced, and the heat transfer on discs is increased considerably at low radii. This is associated with the transition to turbulence in the Ekman layers, as is consistent with the higher boundary layer Reynolds numbers at this condition. The transition is expected to be affected by inlet swirl, which may be of particular interest in engine operating conditions.
  
  \item [4)]
WMLES is recommended for exploring aero engine operating conditions in support of lower order thermal modelling and design in industry, as has been shown here to be feasible. Further investigations on high Reynolds number conditions are needed to confirm and extend the present findings. Exact prediction of Ekman layer transition to turbulence is not expected with WMLES, so more fundamental studies of this are recommended.

\end{itemize}

\section*{Acknowledgments}
Ruonan Wang acknowledges funding from the China Scholarship Council (Grant No.~202004910317) and the University of Surrey. Simulations were conducted on the UK national HPC facility ARCHER2 with support from the UK Turbulence Consortium (Grant No. EP/X035484). The authors thankfully acknowledge the experimental data from the University of Bath, and the discussions with colleagues from the University of Surrey, University of Bath and Rolls-Royce plc. For the purpose of open access, the author has applied a Creative Commons Attribution (CC BY) licence to any Author Accepted Manuscript version arising.



{ \fontsize{9.5pt}{11.8pt}\selectfont

\bibliographystyle{asmeconf}  

\bibliography{draft}
}

\end{document}